\documentclass[aps,prl, superscriptaddress,groupedaddress]{revtex4}  % for review and submission
\usepackage{graphicx}  % needed for figures
\usepackage{dcolumn}   % needed for some tables
\usepackage{bm}        % for math
\usepackage{amssymb}   % for math
\usepackage{xcolor} 
\usepackage{amsmath}
\begin{document}
	\title{Super-Linear Growth Reveals the Allee Effect in Tumors} 
	\author{Youness Azimzade} 
	\email[Email:~]{y_azimzade@ut.ac.ir}
	\affiliation{Department of Physics, University of Tehran, Tehran 14395-547, Iran}
	
	\author{Abbas Ali Saberi} 
	\email[Email:~]{ab.saberi@ut.ac.ir}   
	\affiliation{Department of Physics, University of Tehran, Tehran 14395-547, Iran}
	\affiliation{Institut f\"ur Theoretische Physik, Universitat zu K\"oln, Z\"ulpicher Strasse 77, 50937 K\"oln, Germany}
	
	\author{Robert A. Gatenby}
	\email[Email:~]{Robert.Gatenby@moffitt.org} 
	\affiliation{ Cancer Biology and Evolution Program, Moffitt Cancer Center, Tampa, FL, USA.}
	\affiliation{ Integrated Mathematical Oncology Department, Moffitt Cancer Center, Tampa, FL, USA.}
	\affiliation{Diagnostic Imaging Department, Moffitt Cancer Center, Tampa, FL, USA}
	
	\begin{abstract} 
	\textbf{\textit{Abstract:}}	Integrating experimental data into ecological models plays a central role in understanding biological mechanisms that drive tumor progression where such knowledge   can be used to develop new therapeutic strategies. While the current studies emphasize the role of competition among tumor cells, they fail to explain recently observed super-linear growth dynamics across human tumors. Here we study tumor growth dynamics by developing a model that incorporates evolutionary dynamics inside tumors with tumor-microenvironment interactions. Our results reveal that tumor cells' ability to manipulate the environment and induce angiogenesis drives super-linear growth ---a process compatible with the Allee effect. In light of this understanding, our model suggests that for high-risk tumors that have a higher growth rate, suppressing angiogenesis  can be the appropriate  therapeutic intervention.
	\end{abstract}
	\maketitle
	
	\section{Introduction}
	Tumors comprise a huge number of cells that evolve and expand through a diverse range of interactions between themselves and their environment. Such a complex structure resembles an ecosystem in which different evolutionary processes and ecological interactions exist simultaneously \cite{korolev2014turning}. A relevant, still open problem is to find what processes are more relevant and play the dominant role compared to others \cite{turajlic2019resolving}. Finding the answer to this question is of theoretical interest and is necessary for designing new promising therapies \cite{bozic2013evolutionary, gatenby2020integrating}. Despite numerous research carried on in this field, possible cooperation and its role in the growth and clinical implications have remained less understood, partially due to emphasis on competition between sub-populations in tumor progression \cite{wang2014tumor, parker2020cell, zahir2020characterizing, lakatos2020evolutionary, williams2020measuring}.
	
	Tumors grow in initially healthy tissues with a limited oxygen supply rate. Such a restriction imposes a limit (up to a few mm$^3$) on growing tumors \cite{carmeliet2000angiogenesis}. To expand further, tumors need to increase oxygen supply rate by inducing angiogenesis  ---a process so common and essential that has been introduced as a "Hallmark of Cancer" \cite{hanahan2000hallmarks, hanahan2011hallmarks}. During angiogenesis, each hypoxic cell secretes growth factors that induce angiogenesis, leading to mutualism among tumor cells \cite{archetti2019cooperation}. Cooperation to manipulate the environment in favor of the whole population is analogous to the "Allee Effect" in which individuals cooperate to increase population growth rate \cite{allee1932studies}. While the Allee effect has been suggested to play a role in tumor growth \cite{korolev2014turning, bottger2015emerging, brown2016aggregation, gatenby2019first}, direct evidence for such a role has been scarce \cite{johnson2019cancer}.
	
	Tumor growth dynamics can reveal the biological processes in cellular scale \cite{rodriguez2013tumor, west2019cellular}. In this regards, if all cells can duplicate, for tumor volume (or cell number) one can write: $\frac{dC_T}{dt} \propto C_T^\beta$ with $\beta=1$ which represents an exponential growth \cite{collins1956observation}. For tumors growing in a 3D environment (assuming a well-fed tumor in which cells obey contact inhibition), only cells living on the tumor border can duplicate. Respectively, considering the tumor as a sphere with a radius of $r$, the number of cells on its surface is proportional to $r^2$ and we have $\beta= 2/3$ \cite{mayneord1932law}. Additional limitations, such as oxygen shortage, can decrease this fraction even further and finally set an upper limit on population size \cite{west2001general}. However, recent analysis of different cancers types revealed that tumor growth rate versus volume  increases super-linearly as $\frac{d C_T}{d t} \propto C_T^\beta$ with  $\langle \beta \rangle=1.25$ \cite{Victor2020superlinear}.  
	
	The mechanism behind super-linear growth was attributed to the evolutionary dynamic of competition between cell types. Based on this analogy, once a new cell type with a higher duplication rate emerges, it starts to dominate the tumor and we should observe an increase in tumor growth rate. The continuous emergence of new cell types with higher fitness advantages and their domination increases the tumor growth rate. This process can cause tumors to grow faster than exponentially growing populations and respectively have $\beta>1$ \cite{durrett2010evolutionary, Victor2020superlinear}.  Since the considerably small probability of driver mutation and small fitness advantages they bring are unlikely to have such a drastic effect on $\beta$, the search for the driving mechanism, which is common among different types of cancer, continues. Additionally, tumors with higher growth rates were associated with lower survival times \cite{Victor2020superlinear}, and finding the driver mechanism may have therapeutic implications.  It should be noted that the emergence of new types with higher fitness advantages can increase the growth rate and accelerate invasion \cite{phillips2010evolutionarily,benichou2012front, hallatschek2014acceleration}. However,  it is not clear that if such acceleration is sufficient to significantly increase $\beta$, particularly because even extreme driver mutations can remain unnoticed due to their small effect on overall growth \cite{bozic2019measuring}.  
	
	Here, we developed a multiscale model that includes driver mutations and fitness advantages they bring and cellular oxygen consumption and the ability to induce angiogenesis. Our model suggests that the evolutionary dynamics of competition between sub-populations do not lead to super-linear growth. Instead, super-linear growth emerges during angiogenesis and as a result of the change in the fraction of actively duplicating cells. Thus, super-linear growth dynamics with $\beta>1$ is the direct evidence for mutualism among tumor cells and the Allee effect across human tumors. Additionally, since higher proliferation has come due to more prosperous angiogenesis, suppressing angiogenesis seems to be a proper therapeutic target in high-risk tumors. 
	
	\section{Model}
	 Mathematical modeling has been used to understand and describe different aspects of cancer incidence and evolution over the last century \cite{byrne2010dissecting, altrock2015mathematics}. Interestingly, mathematical description of tumor growth predates our understanding of cancer as a genetic disease \cite{mayneord1932law, schrek1935quantitative}. More recently, mechanistic models have been developed to explain the underlying biological processes that drive tumor growth by including the most fundamental properties of tumor cells such as proliferation and diffusion \cite{anderson2008integrative}. One can consider individual cells and their behavior and study dynamics of a collection of cells where corresponding models are known as individual-based models (IBMs) \cite{metzcar2019review, azimzade2019short}. Alternatively, one can start from a mesoscopic scale and define averaged variables and parameters for a group of cells by taking a mean-field approach (such models are known as continuum).  In this approach, a collection of partial differential equations (PDEs) describe densities of cellular populations and relevant chemicals \cite{swanson2003virtual, mandonnet2003continuous, azimzade2019effect}.  When the number of cells is small, IBMs are preferred due to a more detailed description of individual cells. However, tumors normally contain billions of cells where it is impossible to use IBMs. On such a large scale, continuum models that use PDEs are preferred.  These two approaches can be integrated to generate hybrid models that simulate larger cell numbers with more biological details \cite{rejniak2011hybrid, jimenez2021mesoscopic}. With similar assumptions for population dynamics, IBMs and PDEs may not lead to the same results \cite{azimzade2020invasion} and the connection between the two approaches is not necessarily straightforward. For example, obtaining the corresponding PDE for an IBM can be demanding and lead to various PDEs depending on parameter choices for a particular IBM \cite{johnston2017co}. A common approach to ensure that results are generic and model-independent is to use different continuum and IBM models alongside each other and compare their output, as we will do in this work.  
	
	As our main model, we develop a hybrid model that integrates cells' internal features into external limitations across different scales. We first explain the basic model that rules tumor cells' dynamics and different aspects will be added to the model as we proceed. To include the central role of metabolism in cellular activities \cite{buchakjian2010engine}, we define an internal energy $u$, which is the intensive quantity and is assigned to "each cell". This parameter is regulated by a simple conservation rule as \cite{scalerandi2002inhibition, azimzade2018role}:  
	\begin{equation}
		\frac{\partial u}{\partial t}= -a_u u + b_u n(r,t),
		\label{Eq1}
	\end{equation}
	where $n(r,t)$ is the oxygen density at the position of cell ($(r)$) and time $t$, with $a_u$ and  $b_u$ being positive constants related to energy consumption and accumulation rate.
	
	\begin{figure} 
		\centering
		\includegraphics[width=0.70\linewidth]{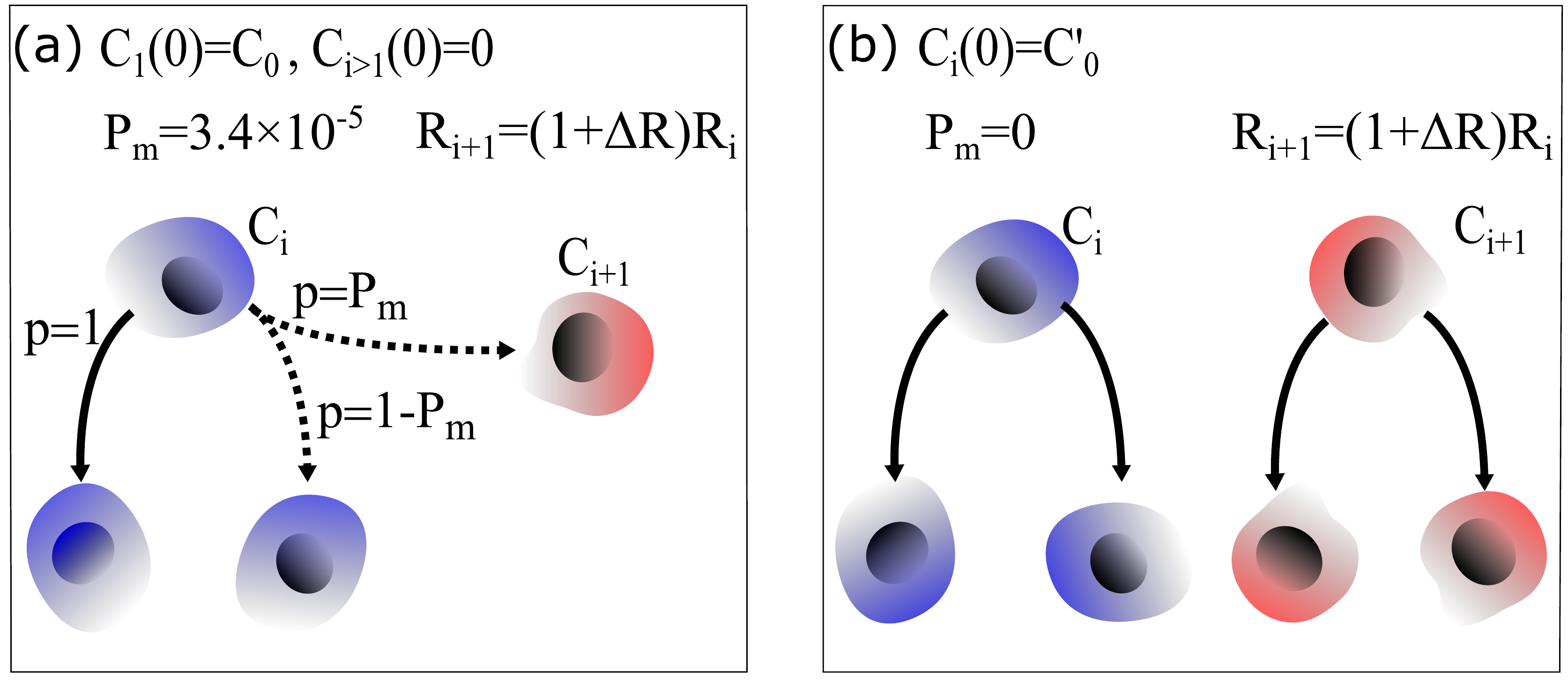} 
		\includegraphics[width=0.39\linewidth]{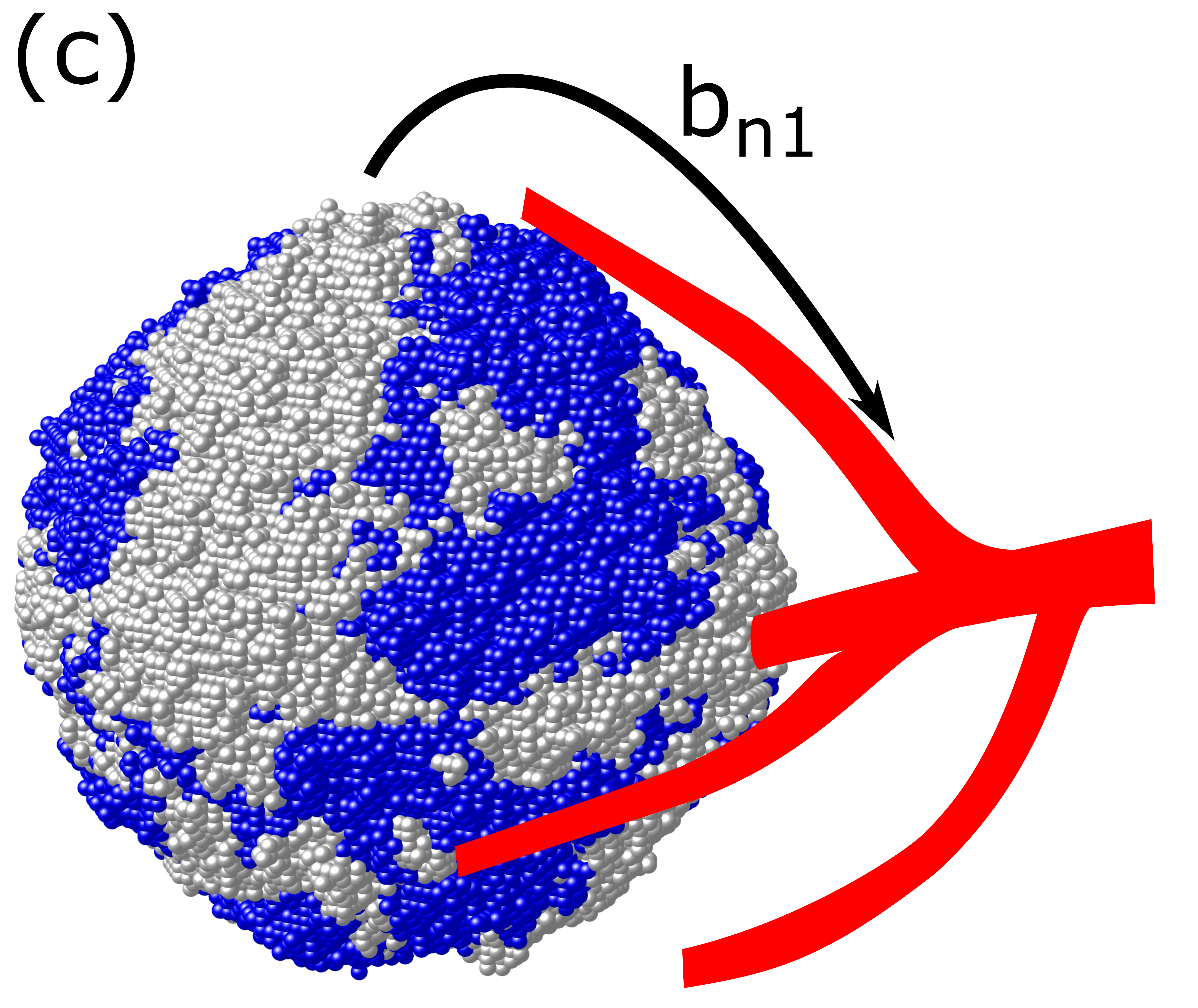}   
		\caption{ (a) Linear (Darwinian) Evolution Model. We start with a single cell type and during each duplication there is chance ($P_m$ which is estimated previously \cite{bozic2010accumulation}) for a driver mutation and appearance of more aggressive cell type. Each driver mutation brings a fitness advantage, $\Delta R$ \cite{bozic2010accumulation}. (b) Punctuated Evolution Model. In this model, different cell types with different fitness advantages live together but there is no transition between cell types. (c) Relation between tumor and oxygen supply system. Tumor cells can increase oxygen supply rate (induce angiogenesis) by $b_{n1}$.}
		\label{FIG1}
	\end{figure}
	
	We consider a 3D lattice with a unit size of 100 $\mu$m which, considering a typical 10 $\mu$m diameter for cancer cells, has the capacity for 1000 cancer cells. Cells live in these units and increase their biomass until they reach the threshold to duplicate. The density of cells of type $i$ at $(r,t)$ is defined by, $C_i(r,t)=\frac{\text{number of cells at}\;(r,t)}{\text{capacity of each site}}$. Tumor cells consume the existing oxygen and the vascular system supplies it for the tissue. As such, the evolution of oxygen is regulated by:
	\begin{eqnarray}
		\frac{\partial n(r,t)}{\partial t}&=&D_n\nabla^2 n(r,t)\\&&- a_n n(r,t)+ b_{n0}- c_n C_t(r,t) n(r,t)\nonumber,
		\label{Eq2}
	\end{eqnarray}
	where $D_n$ is oxygen diffusion constant (equal to $10^{-5}$ cm$^2$/s), $c_n$ is the oxygen consumption rate which is reported to be $6.65\times 10^{-17}$ mol cell$^{-1}$s$^{-1}$ \cite{anderson2005hybrid}, $a_n$ is oxygen decay rate and $b_n$ is oxygen supply rate. $C_t$ is total density of different cell populations and can be obtained as $C_t(r,t)=\sum^g_{i=1} C_{\rm i}(r,t)$ in which $g$ is the number of cell types. 
	
	The evolution of the cells' internal energy, $u$, depends on the local density of oxygen through Eq. (\ref{Eq1}). If cells have access to enough oxygen, $u$ increases to $u_p$ and the cells start to proliferate. Cells that surpass the internal energy threshold follow the standard Fisher-Kolmogorov-Petrovsky-Pishkonov \cite{murray2003mathematical} as: 
	\begin{equation}
		\frac{\partial C(r,t)}{\partial t}=D_c \nabla^2C(x,y,t)+R C(r,t)(1-C(r,t))
		\label{Eq3}
	\end{equation}
	where $D_c$ is the diffusion coefficient  for cells ($\approx 10^{-10}$ cm$^2$/s \cite{anderson2005hybrid}) and $R$ is duplication rate for cells ($\approx 10^{-5}$ s$^{-1}$ \cite{anderson2005hybrid}). For an environment with enough oxygen, cells always have the required internal energy to duplicate with the rate of $R$ and our model automatically becomes the standard FKPP equation.  Such models have been used widely in literature to model tumor growth \cite{swanson2003virtual, mandonnet2003continuous, harpold2007evolution}. They consider two main aspects of cellular behavior: duplication and diffusion. While these parameters are of crucial importance in tumor progression, they are highly sensitive to environmental properties such as oxygen density \cite{hanahan2000hallmarks, hanahan2011hallmarks, whiteside2008tumor}. Our model integrates essential features of cells into environmental limitations by assigning a central role to cells' metabolic states.  We start all simulations with an initial density of cells in the center of a defined environment by setting $C(r,0)=C_0$ for $r<r_0$.  In the following parts, we will add more features to the model to incorporate different possibilities.  
	
	\subsection{Evolutionary Dynamics}
	Here, we consider two main hypotheses that have been used to describe the frequency of sub-populations in tumors,  Linear (Darwinian) Evolution and Punctuated Evolution \cite{wang2014tumor, davis2017tumor, turajlic2019resolving, williams2020measuring}.  The linear evolution suggests that as the tumor grows, new cell types emerge due to mutations and those with higher fitness advantages become dominant, similar to what Darwinian evolution suggests. However, some tumors show high diversity in initial steps, which suggests that they have experienced an abrupt increase in mutations at early stages. A similar process happens in "Punctuated Equilibria," where adaptation occurs in small time intervals and newly adapted individuals rapidly expand out of the niche and through the wider population \cite{gould1972punctuated}. Due to such similarity, punctuated tumor growth evolution is used where different cell types appear at the early stages of tumor growth. Those with higher fitness advantages become dominant as the tumor grows \cite{turajlic2019resolving}. It should be noted that other evolutionary scenarios are possible \cite{davis2017tumor}. Due to a large number of cells and their nonlinear behavior (cell plasticity, for example), the actual dynamics can be more complex \cite{turajlic2019resolving}. Still, we used these two approaches to check if evolutionary dynamics of competition between sup-populations can lead to super-linear growth. 
	
	\subsubsection{Linear Evolution} 
	 As tumors progress, new mutations appear \cite{hanahan2011hallmarks}. However, the detailed mechanism and rate by which mutations happen vary over time and across different cell types. A typical aspect of mutation is that it can be highly correlated. Once a mutation appears in a specif part of DNA, it can lead to instabilities and create a hot spot prone to further mutations \cite{nesta2020hotspots}. On the other hand, mutations can bring different fitness advantages \cite{vermeulen2013defining, mcfarland2014tug}. However, instead of single mutations with huge fitness advantages, it is more accepted that they happen more frequently and gradually lead to variations in cellular features \cite{castro2015mini}. Driver mutations, in general, can lead to higher fitness advantages in different properties of cells. For example, they can lead to higher duplication rates, higher migration rates, or more effective nutrient consumption. However, this process can not go on forever and the differences between fitness advantages are not huge. On the other hand, higher fitness advantages in one aspect are normally associated with loss of other properties \cite{persi2020mutation, hausser2020tumour}. A well-known example of this case is the go-or-grow paradigm \cite{hatzikirou2012go} that suggests cells decide to either migrate or duplicate faster and they can not accomplish both tasks simultaneously. Despite these complexities, simple mathematical models can effectively describe mutations and fitness advantages they bring \cite{waclaw2015spatial, paterson2020mathematical}. As with any random process, mutations are considered to follow a Poison distribution. Thanks to this simplification, the average rate by which driver mutations appear and the corresponding fitness advantages they bring can be calculated. Analysis of experimental results revealed that driver mutations appear with the rate of $3.4\times 10^{-5}$ per duplication and each diver mutation leads to an average 0.4 percent increase in fitness advantage \cite{bozic2010accumulation}. Here we study the effect of driver mutations on duplication and migration rates.   
	
	 In linear evolution, we start with one cell type, $C_1$, by defining the initial condition as $C_1(r, 0) =C_0$ for $r<r_0$ ($C_i(r, 0)=0$ for $i>1$ and $C_1(r,0)=0$ for $r>r_0$).  $C_1$ cells increase their internal energy and duplicate. During each duplication, one daughter cell will have the same type as the parent. The other daughter cell will have the same type with probability of $1- P_m$ and belongs to the next cell type with the probability of $P_m$ with $P_m=3.4\times 10^{-5} $ (  Under this assumption, mutations incidence follows Poison distribution. To include a wider range of possibilities, we include  larger values for $P_m$ up to $10^{-1}$ which is huge compared to any experimental data. ). To take to account such transition between cell types, we need to replace Eq. (\ref{Eq3}) with a new version as below:
	\begin{eqnarray}
		\frac{\partial C_{\rm i}(r,t)}{\partial t}
		= D_c\nabla^2 C_{\rm i}(r,t)\nonumber \\
		+ C_{\rm i}(r,t) \delta_{1i}R_i (1-P_m) [1-C_t(r,t)]  \nonumber \\
		+ P_m (1-\delta_{1i})R_{i-1} C_{\rm i-1}(r,t)[1-C_t(r,t)] 
		\label{Eq4}
	\end{eqnarray}
	where $\delta_{ij}$ denotes the Kronecker delta, i.e., $\delta_{ij}=1$ for $i=j$ and $0$ otherwise, with $1\leq i,j\leq g$. The second term on the right side of Eq. (\ref{Eq4}) is the creation of each type of the cells due to duplication subtracted by those that gain new driver mutations and change their type during duplication. The third term represents the creation of the $i$th type (for $i \ne 1$) of the cells from mutations of the prior type.
	
	Each driver mutation leads to $0.4$ percent increase in fitness advantage \cite{bozic2010accumulation}.  One can take into account increase in fitness advantages due to such mutations  by setting $\Delta R=1- \frac{R_{i+1}}{R_{i}}$ where $R_{i}$ and $R_{i+1}$ are the duplication rates of types $i$ and $i+1$, respectively. Fitness advantages are considered to be small \cite{heide2018reply}. However, to include a wider range in fitness advantages \cite{castro2015mini, vermeulen2013defining, mcfarland2014tug}, we also consider higher values for $\Delta R$ up to $10^{-1}$ which also is huge compared to available experimental data. It should be noted that driver mutations may change other properties in cells as well. For example, similar to other populations, tumor cells may gain higher motility due to driver mutations \cite{benichou2012front}. In this regard, we also considered driver mutations to increase diffusion constant by $\Delta D$.  
	
	\subsubsection{Punctuated Evolution}
	For Punctuated Evolution, we consider different cell types to live together and compete over limited sources. We set their initial values to be equal but their duplication rate to be different as: $R_{i+1}=R_{i} \times (1+\Delta R)$ for $1\leq i \leq g-1$.  Again, to include a wide range of variations in fitness advantages, we will consider different values for $\Delta R$.  Since in this model mutations are considered to be negligible, the evolution of populations follows the equation \ref{Eq4} with $P_m$=0.  We consider 100 cell types (this number does not change the outcome as long as the highest fitness advantage difference remains in a reasonable range --- we have checked up to 800 percent). The frequency of subpopulations with different numbers of mutations is subject to mutation rate and possible fitness advantages they bring \cite{turajlic2019resolving, hausser2020tumour}. We should consider different possibilities for the initial frequency of mutations. For the first case, as the initial condition we set $C_i(r)=C_0$ $C_i(r, 0) =C_0$ for $r<r_0$ ($C_i(r, 0)=0$ for $r>r_0$)  and let the populations evolve based on aforementioned rules. It is widely known for neutrally evolving populations that population number decreases versus mutation frequency, $i$,  as  $1/i$ \cite{williams2016identification}. We can implement such a scenario by considering $C_i (r,0) \propto C_0/i$ as the initial condition for $r<r_0$. These two evolutionary models will be used to check if evolutionary dynamics of competition between different cell types can lead to the emergence of super-linear growth. 
	
	\subsection{Angiogenesis}
 Angiogenesis occurs when starving cells start to secrete chemicals such as  Vessel Endothelial Growth Factor (VEGF) to induce endothelial cell duplication and formation of vessels. Endothelial cells in response provide starving cells with nutrients. A precise model can take into account the density of endothelial cells, the density of starving cells, the rate by which chemicals are secreted/up taken, as well as their temporal and spatial evolution \cite{scalerandi2002inhibition, phillips2020hybrid}. However, such an approach leads to an excessive number of parameters \cite{phillips2020hybrid, zheng2018mathematical}. For the sake of simplicity, we take a different approach. Rationally, one can assume the angiogenesis (and subsequent increase in oxygen supply rate) to be a monotonic function of tumor volume \cite{d2007rapidly}. Additionally, oxygen supply rate can be a monotonic function of vascular density. Putting together, a single parameter can effectively describe how angiogenesis affects oxygen supply rate. As such, oxygen evolution (previously described by Eq. (\ref{Eq2})) will have an additional term as: 
		\begin{eqnarray}
			\frac{\partial n(r,t)}{\partial t}&=&D_n\nabla^2 n(r,t)\\&& - a_n n(r,t)+ b_{n0}+ b_{n1} C_T - c_n C_t(r,t) n(r,t)\nonumber,
			\label{Eq6}
		\end{eqnarray}
		where $b_{n1}$ (in mol cell$^{-1}$ s$^{-1}$mm$^{-3}$) depicts the increase in oxygen supply as a result of angiogenesis.  This approach is conceptually in line with the previous studies on the dynamics of vasculature \cite{d2007rapidly, rieger2015integrative, tracqui2009biophysical, xu2016mathematical}. However, contrary to what one may naively assume, angiogenesis is not a linear process. A  vascular tumor may fail to grow larger due to failure in inducing more angiogenesis, a process known as post-vascular dormancy \cite{hahnfeldt1999tumor, naumov2008tumor}. 
	
 \subsection{Solving the model} 
 We have two kinds of inter-related variables that evolve over time, oxygen density and cellular densities.  Oxygen density is constantly changing due to supply and consumption by cells. On the other hand, cellular density evolves if cells have accumulated enough internal energy. Respectively, equations that describe these variables are not coupled as one may see in conventional approaches and allows us to solve equations separately. At each time step, we solve oxygen density using Forward Time Centered Space (FTCS). For simplicity, we explain the approach for the 1D version of Eq. (2) and extension to higher dimensions is straightforward. 
		Discretization of the reaction terms (consumption, decay and supply) is straightforward. For Laplacian-like term in a uniformly-space grid with a set of grid points $x_i=i\delta$ ($i=0\l,\pm 1\;,\pm 2,\cdots$), let $n(x_i,t)=n_i(t)$, and use central difference-discretization, $\mbox{\boldmath$\nabla$}n\approx 
		(n_{i+1}-n_{i-1})/(2\delta)$, where  $j$ belongs to nearest neighbors of $i$ and  $\delta$ is the distance between $i$ and $i+1$, we obtain: 
		\begin{equation}
			\frac{\partial n_i}{\partial t}=D_{n}(n_i-n_{i-1})+D_{n}(n_i-n_{i+1})- a_n n_i(t)+ b_{n0}- c_n C_t(x_i,t) n_i(t) \;.
	\end{equation}  
	
	  As mentioned before, we set unit size (and respectively distance between two neighboring grid points)  at $100\;\mu$m.  For each grid site we solve for $n$ by writing: 
		\begin{displaymath}
			n_i(t+\Delta t)=n_i(t)+ \Delta t \sum_{j\in\{i\}}D_{n}[n_j(t)-n_i(t)] +\Delta t \big(-a_n n_i(t)+b_{n0}-c_n C_t(x_i,t) n_i(t) \big)\;,
		\end{displaymath}
		where $\Delta t$ is the time step. FTCS method  typically converges to the solution and is stable. Stability requires that 
		$D\Delta t/(4\delta^2)<0.5$, suggesting  $\Delta t \sim 10 s$. As the initial condition for oxygen, we set $n(r,t=0)= b_{n0} /a_n$.  For cellular densities that surpass the energy threshold, the corresponding FKPP equation was solved using FTCS method \cite{azimzade2019effect} and codes are available at \url{https://gitlab.com/YounessAzimzade/superlinearallee}.

 \subsection{Parameters and Variables} 
 To give a comprehensive view to readers, here we present all variables that we have defined (Table I)  and parameters we have used (Table II) in our model.   \\
	
	\begin{table}[h]
		\centering
		\caption{ Variables defined in the model and their definitions and values. }
		\setlength{\arrayrulewidth}{0.1mm}
		\setlength{\tabcolsep}{15pt}
		\renewcommand{\arraystretch}{0.8}	
		\begin{tabular}{|p{1.0cm}|p{10cm}|}
			\hline
			Variable & Definition   \\
			\hline
			$u$  & Internal energy of each cell \\ 
			\hline
			$n$  & Oxygen density \\ 
			\hline 
			$C$  & Cellular density defined as $C_i(r,t)=\frac{\text{number of cells  at}\;(r,t)}{\text{capacity of each site (1000)}}$ \\ 
			\hline
			$C_i$  & Density of type $i$  \\ 
			\hline
			$C_t$  & Total  cellular density defined by $C_t(r,t) =\sum^g_{i=1} C_{\rm i}(r,t)$  \\ 
			\hline
			$C_T$  & Tumor volume  $C_T(t)=\int C_t(r,t)dV$  \\ 
			\hline
			$\frac{d C_T(t)}{d t} $ & Tumor growth rate \\ 
			\hline 
		\end{tabular}
	\end{table} 
	\begin{table}[h]
		\centering
		\caption{  Parameters used in the model and their definitions and values. }
		\setlength{\arrayrulewidth}{0.1mm}
		\setlength{\tabcolsep}{15pt}
		\renewcommand{\arraystretch}{0.8}	
		\begin{tabular}{|p{1.3cm}|p{7cm}|p{4.9cm}|}
			\hline
			Parameter & Definition & Value \\
			\hline 
			$a_u$   & Energy consumption rate & $0.1$ (estimated) \\	
			\hline
			$b_u$   & Energy accumulation rate & $1$ (estimated)\\	
			\hline
			$u_{p}$   & Proliferation energy threshold & $8$  (estimated)\\
			\hline
			$R$   & Cellular duplication rate & $0.05$  \cite{haass2014real} \\	
			\hline
			$D_C$   & Cell  diffusion  constant &  $10^{-10}$cm$^2$/s  \cite{anderson2005hybrid, bray2001cell} \\ 	
			\hline
			$D_n$   &  Oxygen diffusion constant in tissue & $10^{-5}$ cm$^2$/s \cite{anderson2005hybrid} \\  
			\hline
			$c_n$   & Oxygen consumption rate & $6.65\times 10^{-17}$ mol cell$^{-1}$s$^{-1}$ \cite{anderson2005hybrid}    \\ 
			\hline
			$b_{n0}$   & Oxygen supply rate &  variable - estimated    \\ 
			\hline
			$b_{n1}$   & Increase in oxygen supply rate & variable - estimated    \\ 
			\hline
			$a_{n}$   & Oxygen decay rate &  10$^{-4}$ - estimated    \\ 
			\hline
			$P_{m}$   & Driver mutation rate &  $3.4\times 10^{-5}$ per division  \cite{bozic2010accumulation}    \\ 
			\hline
			$\Delta R$ & Fitness Advantage of each driver mutation &  $4\times 10^{-3}$  \cite{bozic2010accumulation}    \\ 
			\hline
		\end{tabular}
	\end{table} 
	
	\section{Results } 
	We first study the basic model and then will analyze how adding new aspects affect the growth dynamics. To quantify growth dynamics, we measure tumor volume, $C_T(t)=\int C_t(r,t)dV$ and the accumulated duplication of all cells, $\frac{dC_T(t)}{d t}=\int\frac{\partial C_t(r,t)}{\partial t} dV$, where the latter gives the growth rate for whole tumor. We first check two possibilities: (a) Where the oxygen supply rate is high  ($b_{n0}=10^{-4}$ mol s$^{-1}$mm$^{-3}$) and all cells will obtain the requisite oxygen to duplicate. (b) Where the oxygen supply rate is low ($b_{n0}=5 \times 10^{-6}$ mol s$^{-1}$mm$^{-3}$) and its shortage halt the growth when tumor volume approaches $\sim$2 mm$^3$ --- about the maximum size tumors can grow without angiogenesis.
	\begin{figure} [t]
		\centering
		\includegraphics[width=0.48\linewidth]{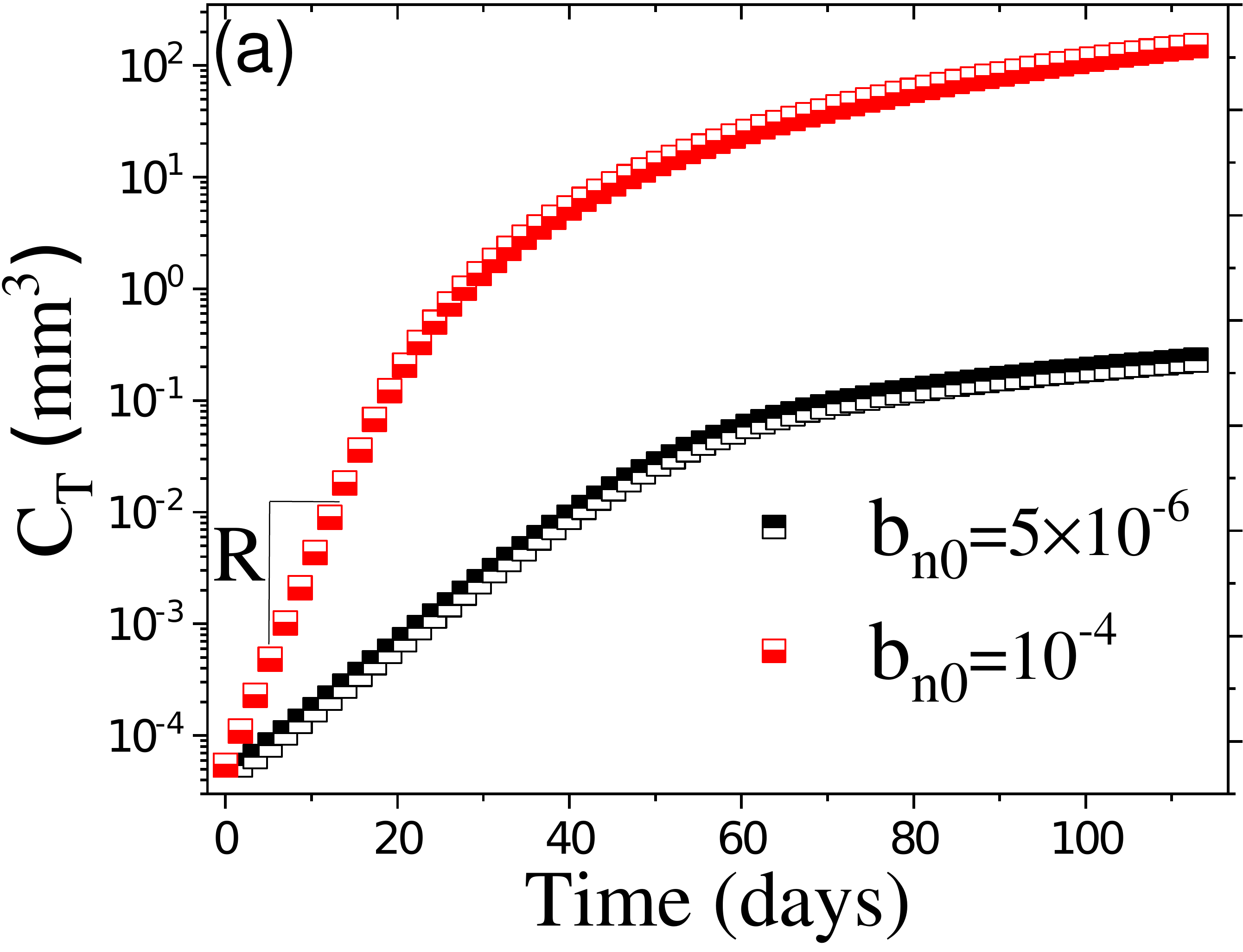} 
		\includegraphics[width=0.48\linewidth]{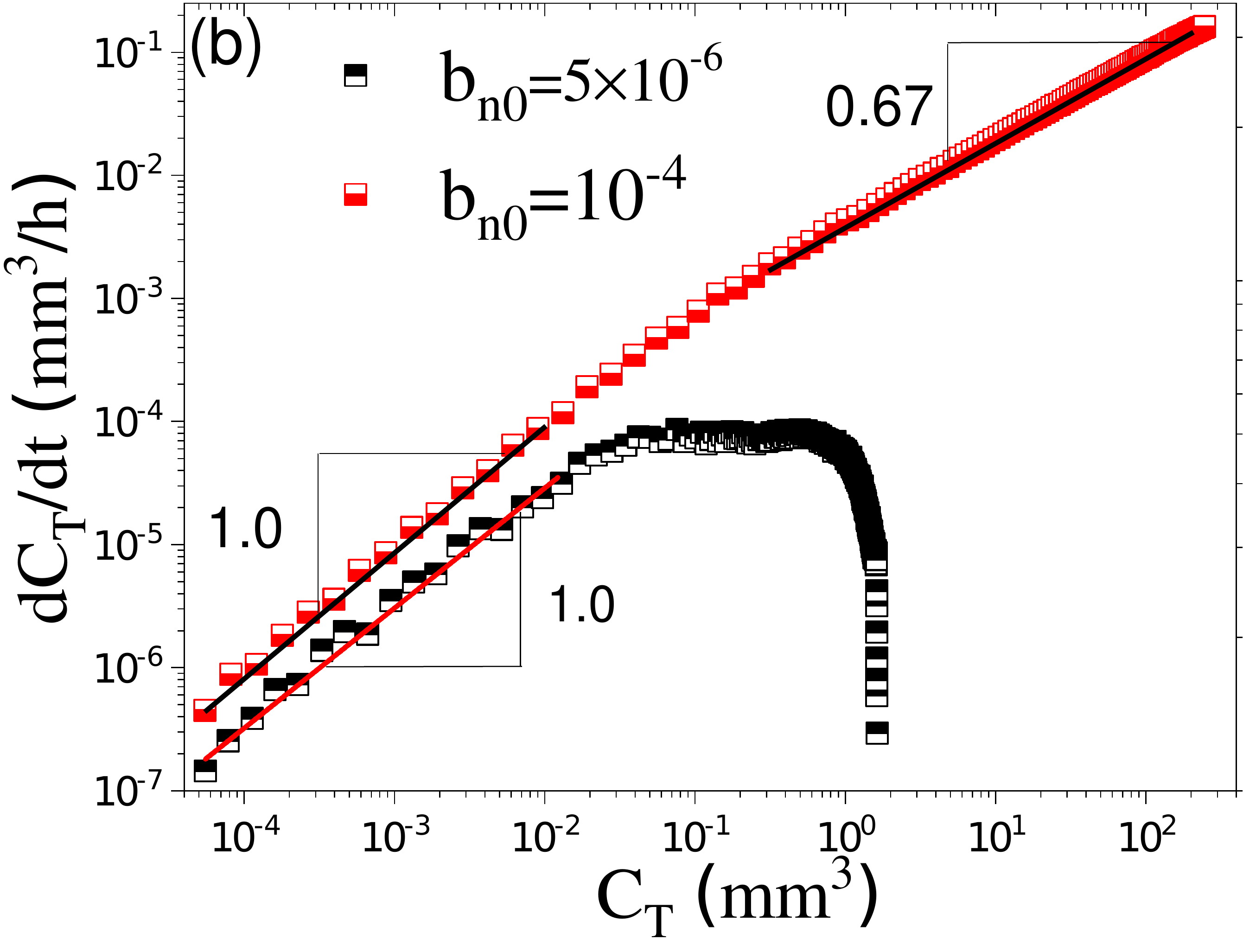} 
		\caption{(a) Volume versus time for different oxygen supply rate ($b_{n0}$ in mol s$^{-1}$mm$^{-3}$) in the half-logarithmic diagram. Both high and low oxygen supply initially show exponential growth. (b) Growth rate versus volume for different values of oxygen supply. Oxygen shortage halts growth and induces smaller values for $\beta$. Limited oxygen supply forces an upper limit where the population size can not surpass it.}
		\label{FIG2}
	\end{figure}
	
	For high oxygen supply rate, our model converges to a traditional 3D FKPP equation. As such, initially all cells can duplicate and we observe an exponential growth. Thus, one has $\frac{d C_T(t)}{d t} \sim C_T(t)$ and tumor volume grows as $C_T(t)=C_T(0) e^{Rt}$ (see early times dynamics in FIG. \ref{FIG2} (a) and FIG. \ref{FIG2} (b)  where  we have $\beta=1$).  For a considerably large tumor, only cells that are living on the surface will have a chance to duplicate and fraction of actively duplication cells for a spherical tumor of radius  $r$ becomes proportional to $ r^2$. With minor simplification,  we have: $\frac{d C_T(t)}{d t} \sim r^2$. Thus, in the log-log diagram we have the $\frac{d C_T(t)}{d t} \sim C_T(t)^{\beta}$ with $\beta=2/3$, as FIG. \ref{FIG2} (b) shows. 
	
	In the second case, which is more realistic, considering a wide range of evidence for hypoxia, the oxygen supply rate is lower and cells duplicate slower. Yet, as FIG. \ref{FIG2} (a) shows, tumor grows exponentially at earlier stages and we have $\beta=1$ (see FIG. \ref{FIG2} (b)). As the tumor grows larger, more and more cells face a nutrient shortage, and duplication becomes slower. Limited oxygen supply imposes an upper limit on population size. When tumor size reaches the limit, growth stops and $ \frac{d C_T(t)}{d t}$ approaches zero.  
	
	\subsection{Evolutionary Dynamics}
	Here we check if evolutionary processes can lead to the emergence of super-linear dynamics. Theoretically, emergence of new cell lines with higher fitness advantages can accelerate the growth \cite{durrett2010evolutionary, Victor2020superlinear} (see FIG. \ref{FIG3} (a)).
	
	\subsubsection{Linear Evolution}
\begin{figure} 
		\centering
		\includegraphics[width=0.55\linewidth]{FIG3a.pdf}
		\includegraphics[width=0.48\linewidth]{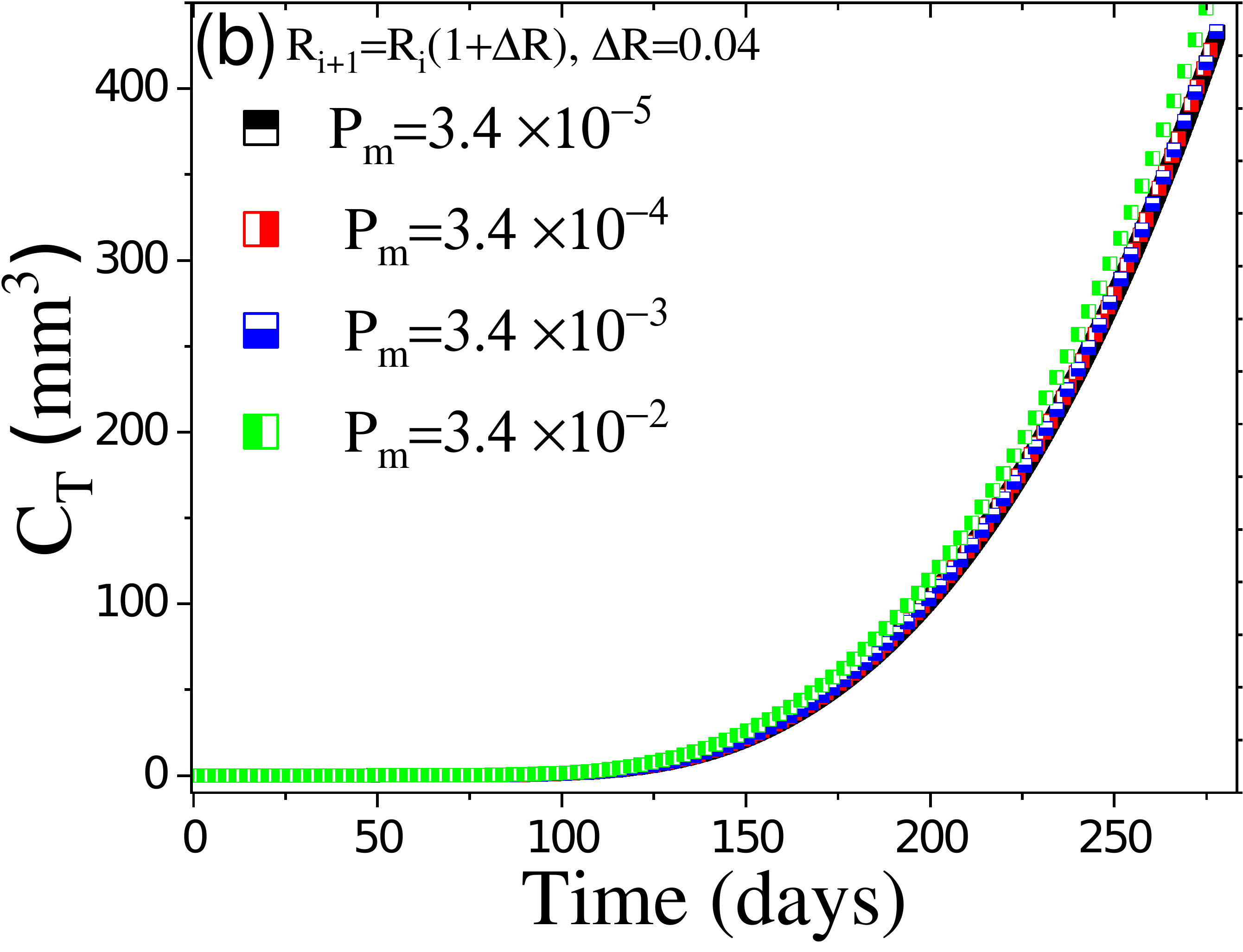} 
		\includegraphics[width=0.48\linewidth]{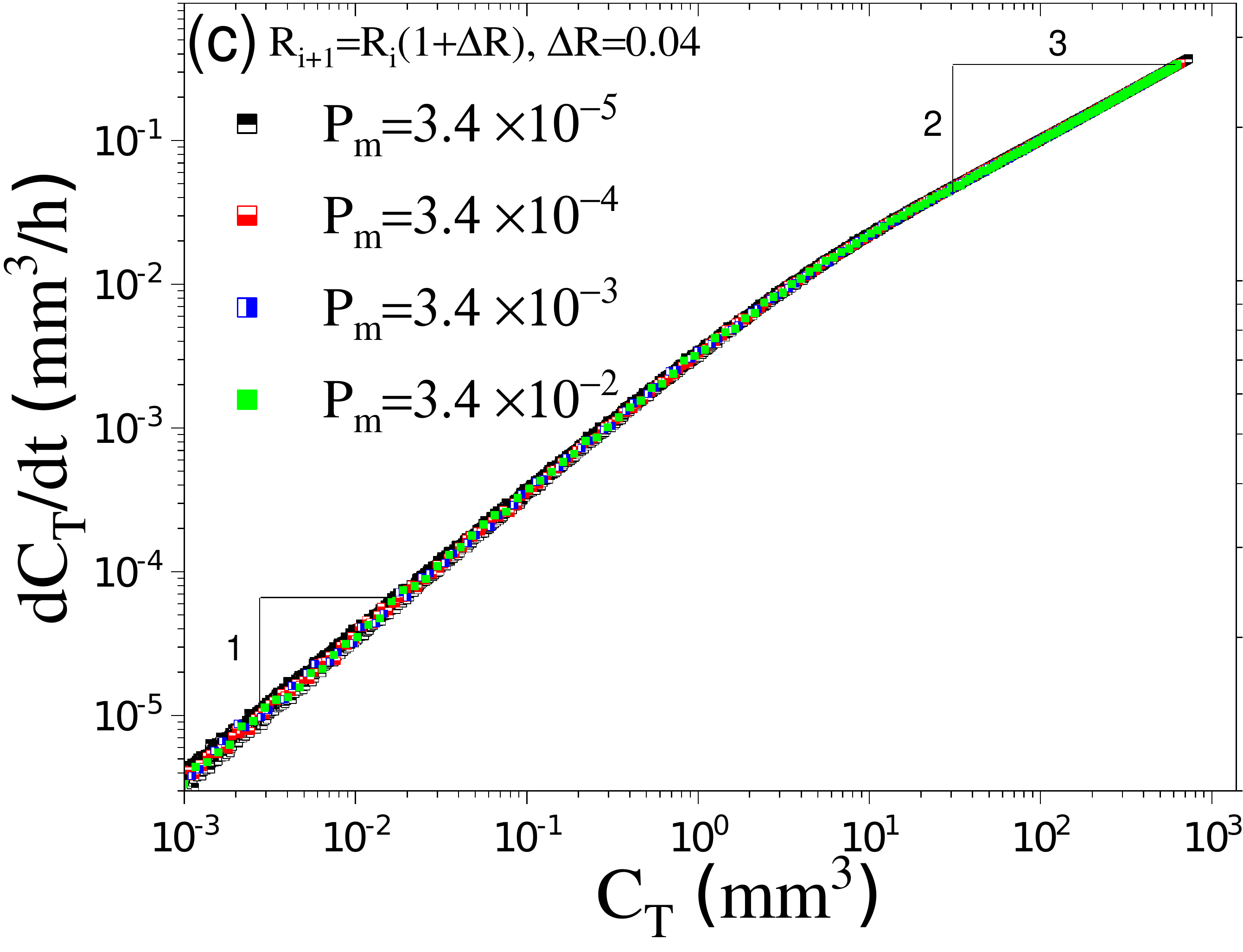} 
		\includegraphics[width=0.48\linewidth]{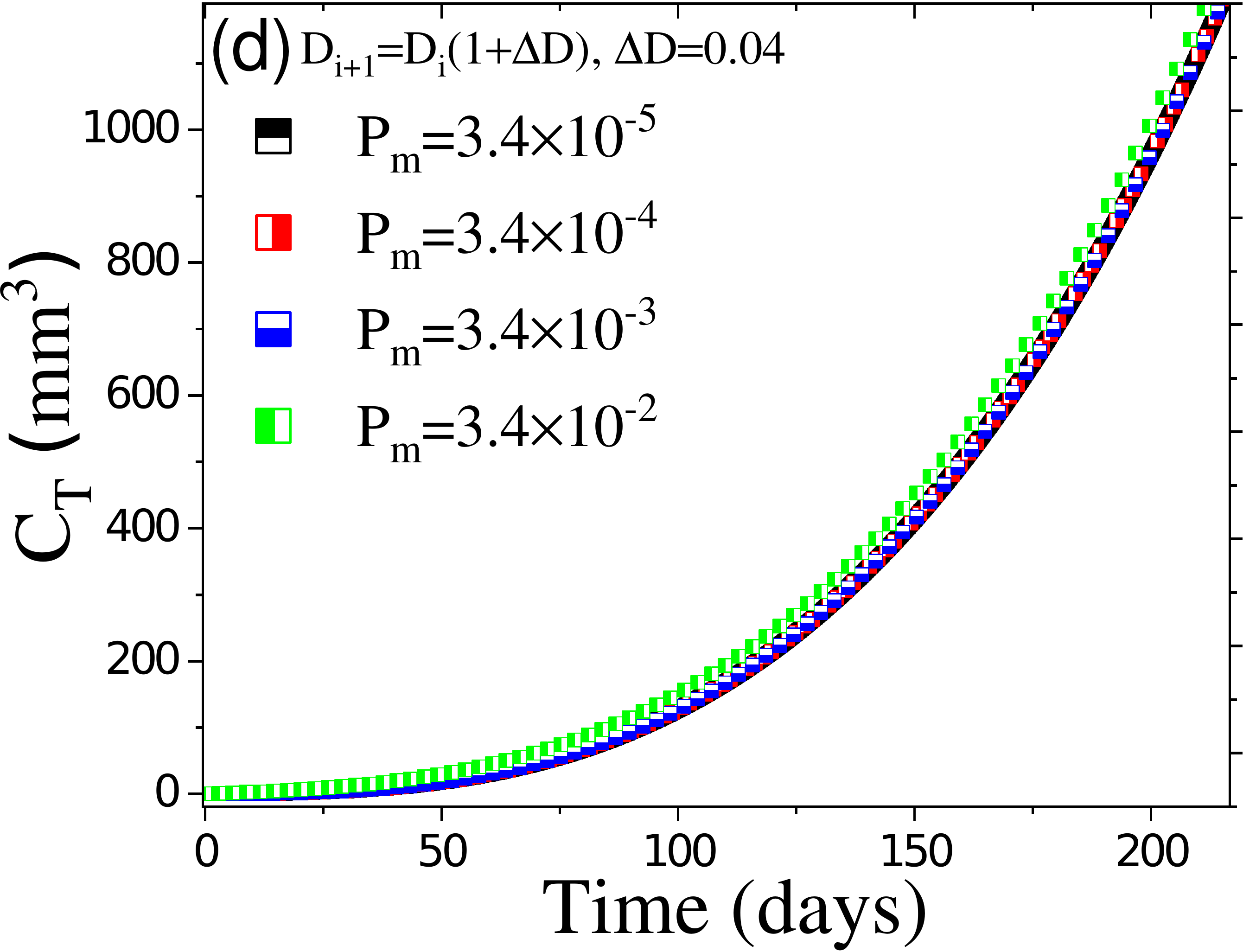} 
		\includegraphics[width=0.48\linewidth]{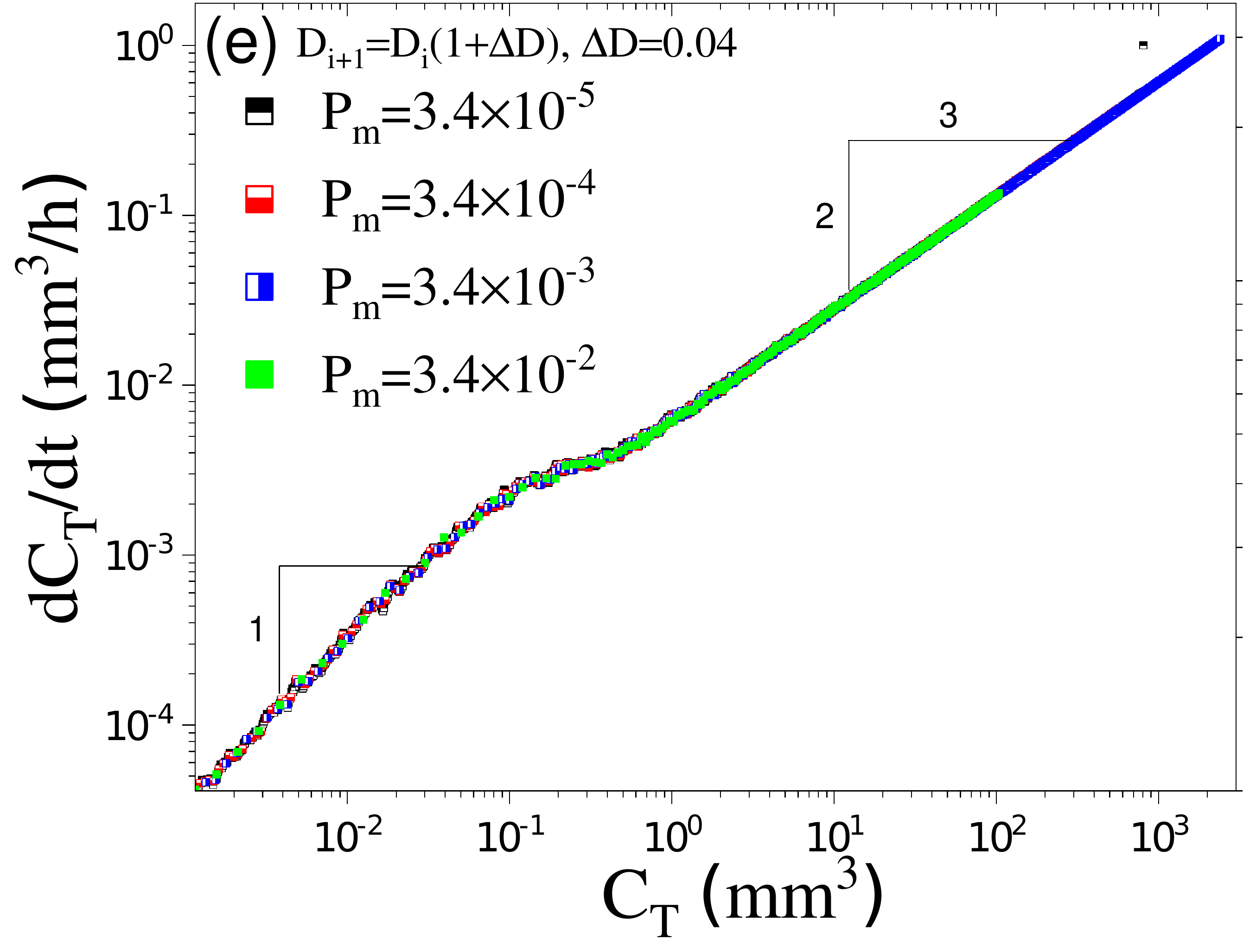} 
		\caption{(a) Emergence of cell types with higher proliferation rate can lead to accelerated growth which is faster than exponential growth and exhibit $\beta>1$ \cite{durrett2010evolutionary, Victor2020superlinear}.  (b) Tumor volume ($C_T$) versus time for different values of $P_m$ and $\Delta R=0.04$. (c) Growth rate versus tumor volume for the same tumors.  (d) Tumor volume versus time for $\Delta D=0.04$ and different values of $P_m$. (e) Growth rate versus tumor size for the same tumors.  Analysis of growth dynamics for linear evolution suggests that emergence of new sub-populations with higher fitness advantage is not able to impose a drastic change on $\beta$.}
		\label{FIG3}
	\end{figure}

	We start the simulation with a single cell type and allow new cell types to arise due to mutations.  We include different values for  $\Delta R$ and $P_m$ and assume that driver mutations increase duplication rate. As FIG. \ref{FIG3} (b) shows tumor volume versus time, higher mutation rate (fitness advantage) lead to faster growth. However, this change is very small for experimentally relevant parameters. On the other hand, we need to check if this acceleration in growth is big enough to cause super-linear growth.  As an analysis of growth dynamics in  FIG. \ref{FIG3} (c) shows, even considerably large values of $\Delta R$ and $P_m$ do not increase $\beta$.  
	
	   Similar to previous part, we consider driver mutations to occur with the rate of $P_m$ and each driver mutation diffusion constant  by the factor of $\Delta D$ as $D_{i+1}=D_i (1+\Delta D)$. As FIG. \ref{FIG3} (d) shows, even considerably huge rates for driver mutations ($P_m=3.4 \times 10^{-2}$) and high fitness advantages ($\Delta D=0.04$) increases growth rate slightly. More importantly, such an increase in growth rate does not affect growth rate and as FIG. \ref{FIG3} (e) shows the growth rate remains almost the same. 
	
	 Emerging new cell types with higher duplication (migration) rate increase invasion velocity and accelerates the growth, but it does not necessarily lead to $\beta>1$. To have large values of $\beta$, the newly emerged types should take over the population in short time intervals. New cell types have small fitness advantages to overcome existing populations in short times. Additionally, for compactly growing populations, the clonal sweep (domination of one type) is a slow process \cite{hallatschek2014acceleration}. That is why we do not observe the fast domination of new types that hypothetically could lead to super-linear growth. Similar results on the negligible effect of driver mutations on growth rate have been reported previously with an entirely different model \cite{waclaw2015spatial}. Due to such small effects, it is difficult to detect driver mutations \cite{bozic2019measuring}.   
	
	\subsubsection{Punctuated Evolution} 
\begin{figure} 
		\centering
		\includegraphics[width=0.48\linewidth]{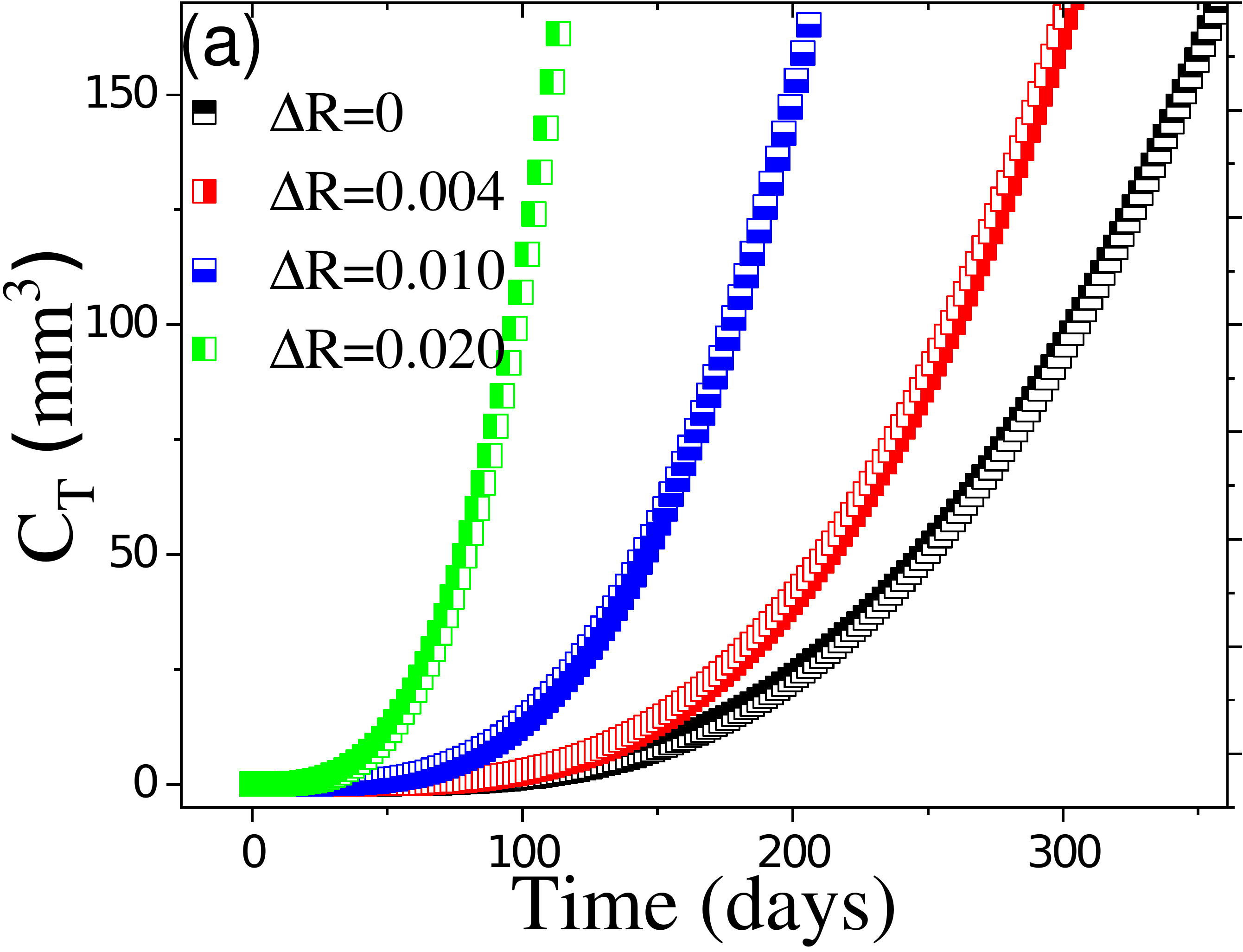}  
		\includegraphics[width=0.48\linewidth]{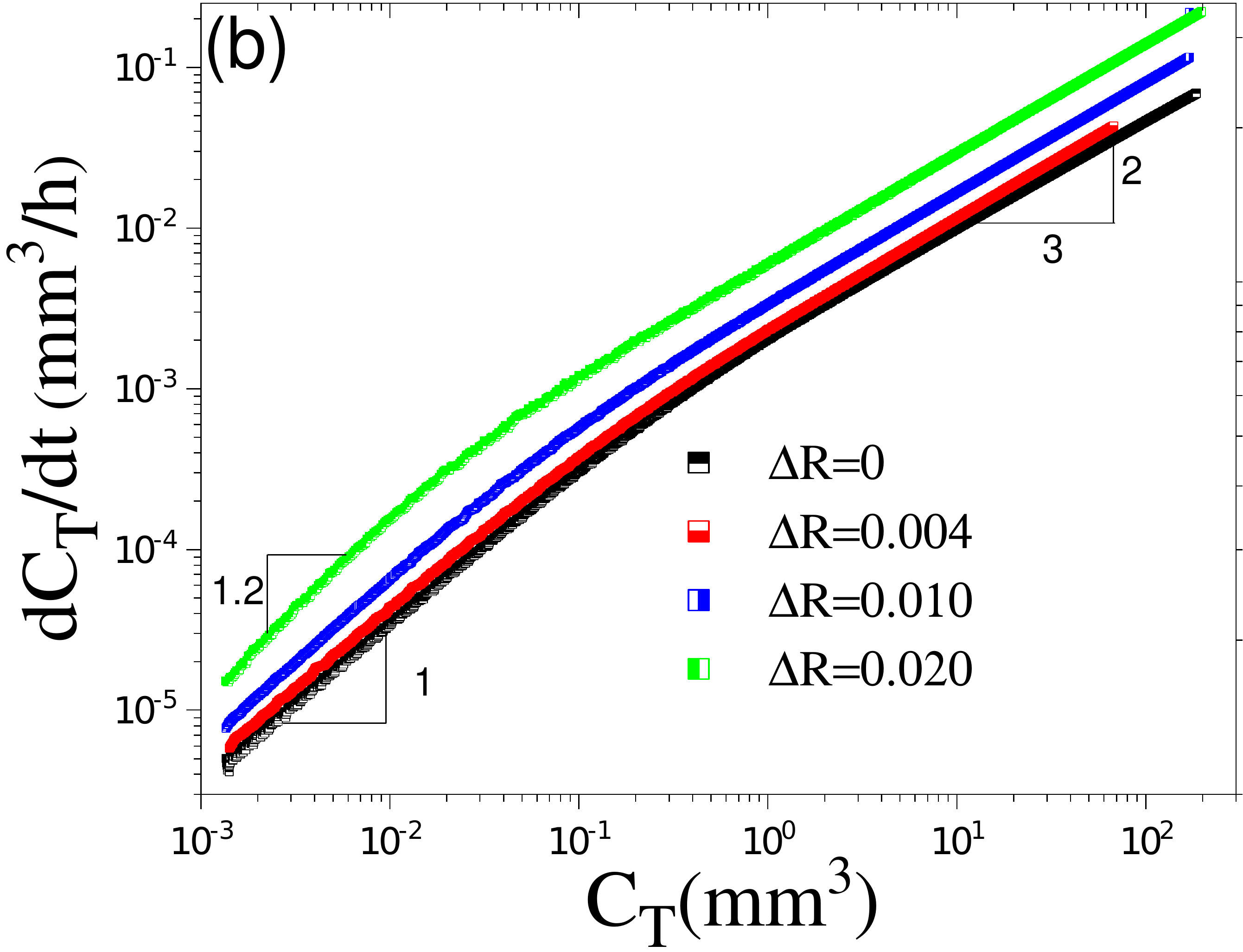} 
		\caption{  (a) Volume versus time for Punctuated evolution with uniform distribution for cellular densities ($C_i(0)= C_0$). (b) Growth rate versus tumor size for the same tumors.  Experimentally relevant values of differences in cellular fitness does not affect growth dynamics. Large  values of $\Delta R$ lead to huge differences in duplication rates and leads to $\beta>1$ for very small tumor sizes where the growth is exponential for $\Delta R=0$. Even such  a huge differences in duplication rates do not change $\beta$ in larger and clinically relevant sizes.   }
		\label{FIG4}
	\end{figure}

 Depending on the value of $\Delta R$, fitness advantage varies across cell types with a different number of mutations. With $\Delta R=0.004$, $\Delta R=0.01$ and $\Delta R=0.02$, we have $\frac{R_{100} }{R_1}=1.48$,  $\frac{R_{100}}{R_1}=2.68$ and $\frac{R_{100}}{R_1}=7.24$, respectively. Such differences are huge compared to experimental results (10-20 percent difference \cite{heide2018reply}), but we include them to check for extreme cases. We start simulations with uniform distribution across cell types ( by $C_i (0)= C_0$)  and allow cell populations to evolve based on the aforementioned rules and Eq. \ref{Eq4} with $P_m=0$. As FIG. \ref{FIG4} (a) shows, increasing $\Delta R$ leads to considerably faster growth. Doubling times for small tumor sizes ($C_T < 0.1 $ mm$^3$ where the growth is exponential) with $\Delta R=0$, $\Delta R=0.004$, $\Delta R=0.01$ and $\Delta R=0.02$ are of 21,  17.5, 11  and 6 days, respectively. For small tumor sizes where the growth is exponential, growth can exceed exponential growth as more fit types start to dominate the tumor, as suggested by \cite{durrett2010evolutionary} and FIG. \ref{FIG3} (a) shows.  This process can lead to $\beta>1$ in small tumor sizes, but it does not lead to super-linear growth in larger, clinically relevant, tumor sizes as FIG. \ref{FIG4} (b) shows.  Two aspects of this process needs more clarifications: First, for experimentally relevant parameters ($\Delta R \le 0.01$) even small tumors do not show super-linear growth ( $\Delta R=0.004$ at FIG. \ref{FIG4} (b)). Super linear growth emerges for $\Delta R \sim  0.02$ where the $C_100$ can duplicate 7.25 times faster than $C_1$. Clearly such a huge difference in tumor cells proliferation rate has not been reported yet and seems impossible considering limitations that cells face in their way to significantly increase their fitness advantage \cite{hausser2020tumour}. More importantly, even such a huge difference can only lead to super-linear growth in small tumor sizes. As FIG. \ref{FIG4} (b) shows, for larger and clinically relevant sizes ($C_T \sim 10^3$ mm$^3$) where super-linear growth has been observed \cite{Victor2020superlinear}, even these huge differences do not lead to super linear growth. 
	
\begin{figure} 
		\centering
		\includegraphics[width=0.48\linewidth]{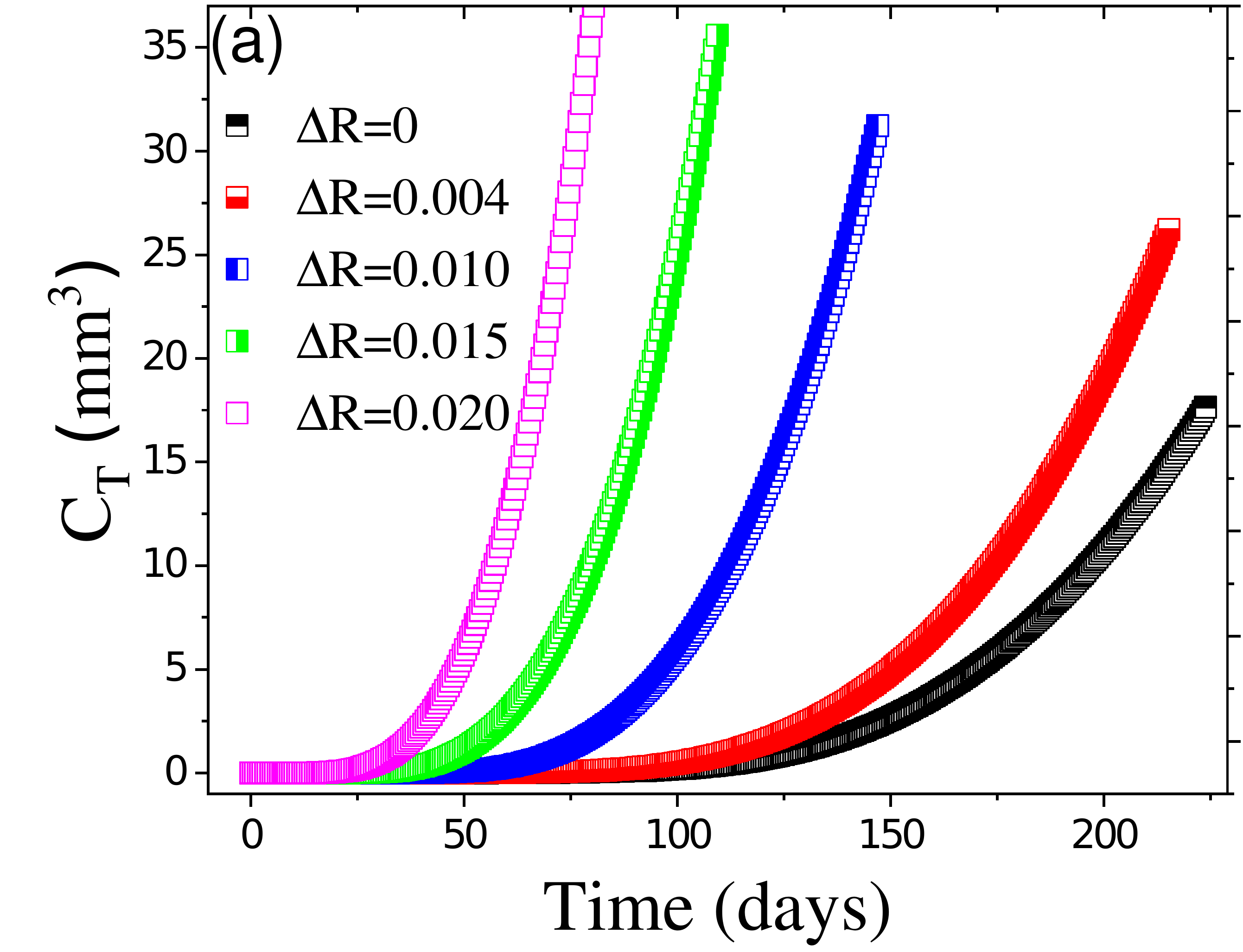}  
		\includegraphics[width=0.48\linewidth]{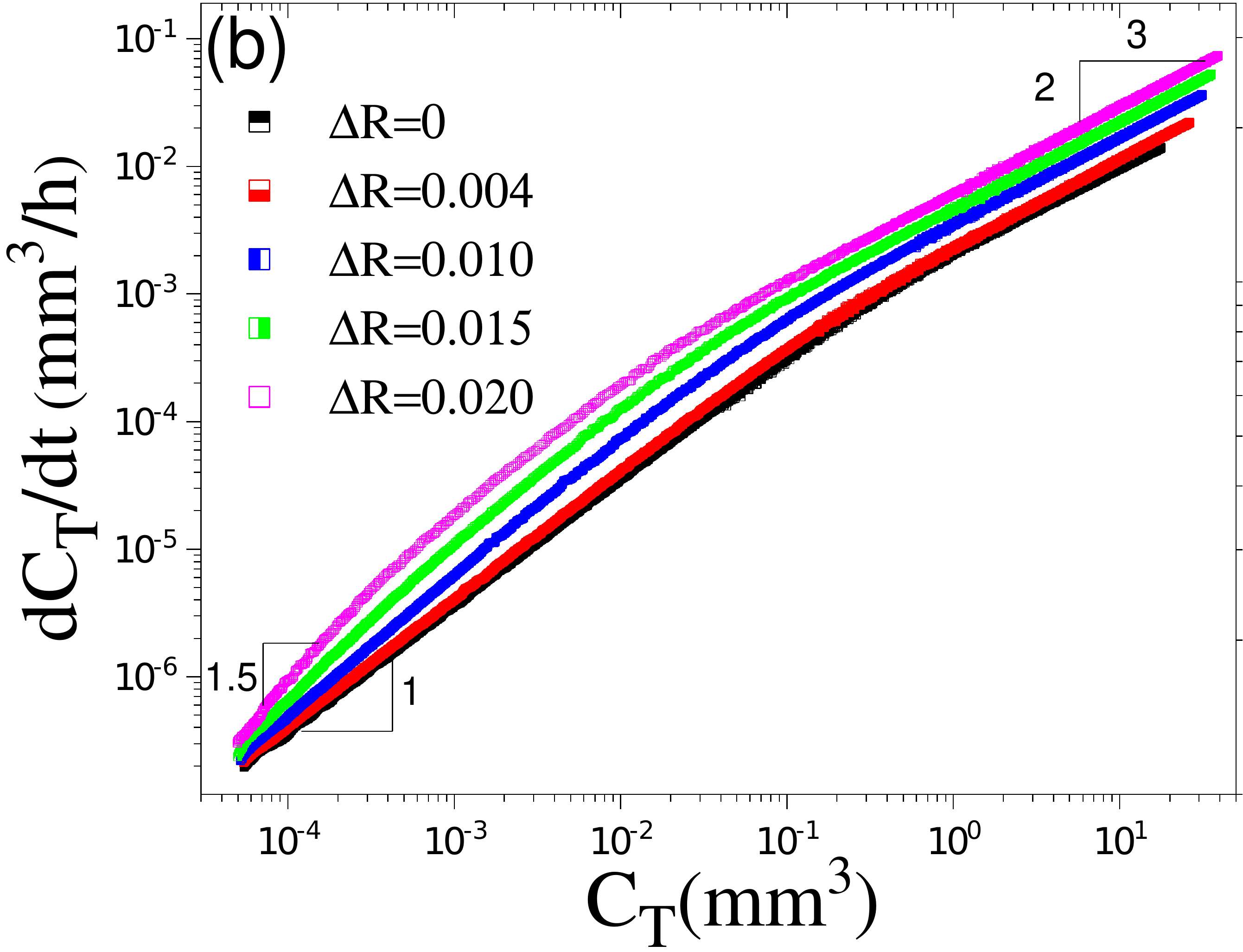}
		\includegraphics[width=0.48\linewidth]{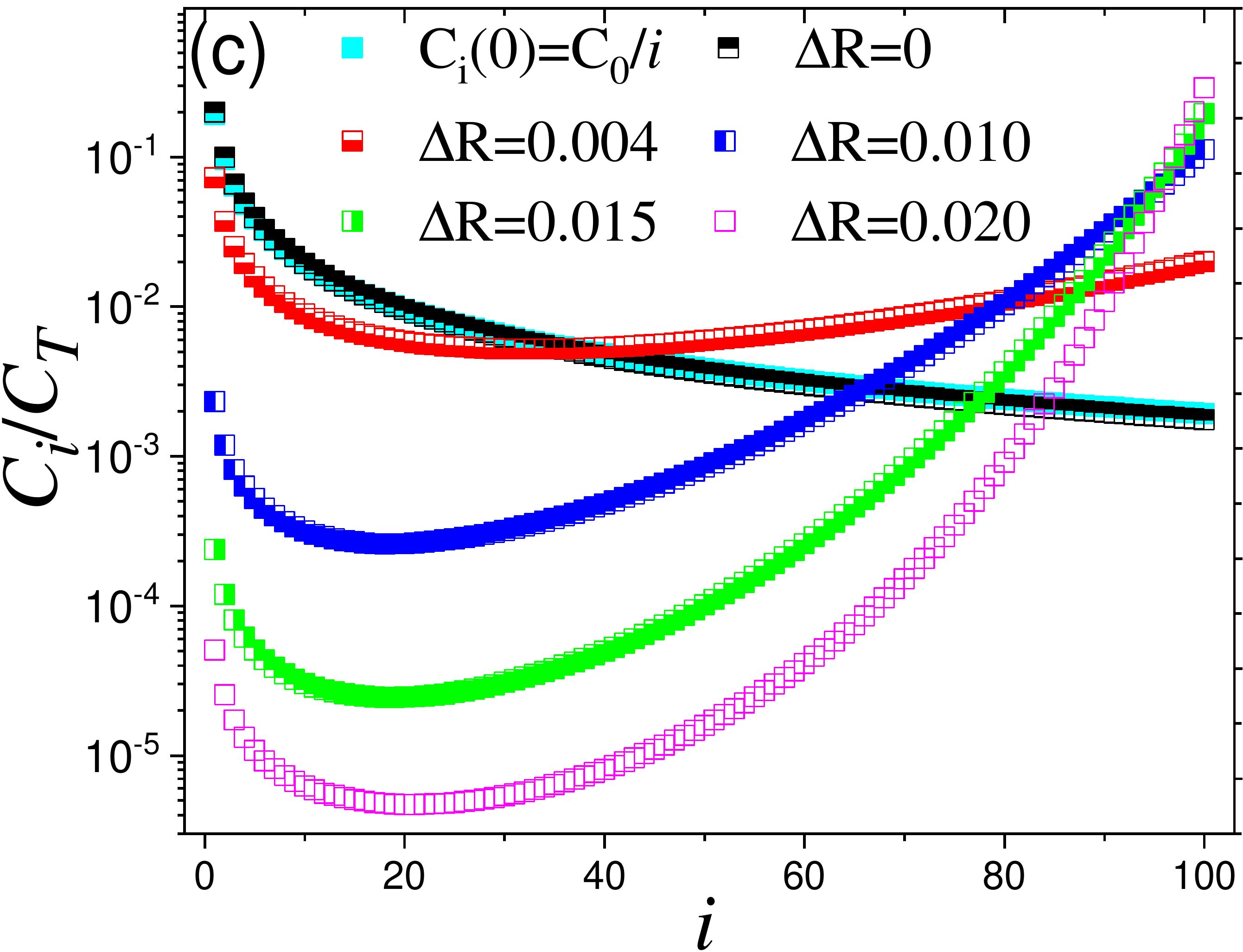} 
		\caption{ (a) Volume versus time for different values of $\Delta R$ and with $C_i(0)=C_0/i$ as the initial condition. (b) Growth rate versus tumor size for the same tumors.  While in smaller sizes we can observe $\beta>1$, larger tumors exhibit $\beta=2/3$. (c) Fraction of each cell type for tumors that have been growing for 90 days. In tumors with high $\Delta R $, cells with higher duplication rate are able to take over whole population which leads to $\beta>1$. later on, such tumor exhibit $\beta=2/3$. }
		\label{FIG5}
	\end{figure}

  For non-uniform density of cell types at initial point ($C_i (0)= C_0/i$), we again check for growth dynamics. Similar to the previous part, larger values of $\Delta R$ lead to faster growth (see FIG. \ref{FIG5} (a)). Interestingly, non-uniform distribution leads to larger values of $\beta$ compared to uniform distribution (see FIG. \ref{FIG5} (b)). However, even with such high values for $\beta$ at initial times, $\beta$ approaches 2/3 as the tumors grow larger. As FIG. \ref{FIG5} (c) shows, for $\Delta R>0$, the composition of tumor changes and the frequency of sub-populations with higher duplication rate increases. Depending on the value of $\Delta R$, two main scenarios can happen: for larger values of $\Delta R$, population(s) with higher duplication rate take over the whole population at initial times and once the tumor becomes slightly larger, the growth approaches to growth with $\beta=2/3$. On the other hand, if $\Delta$ is smaller, populations with higher duplication rates can not take over the whole tumor in short time intervals and if finally, they do, the process is slow and does not affect $\beta$. To cross-check our results, we developed an individual-based model for exponential growth with the same mutation rates and fitness advantage variation and the results confirm our findings here \cite{supplemtal}. 
	
	\subsection{Angiogenesis}   
	As mentioned, tumors normally face an oxygen shortage in the early stages, and considering a low oxygen supply rate, for example, $b_{n0}=5\times 10^{-6}$ mol s$^{-1}$mm$^{-3}$, is rational. With no angiogenesis, such a tumor can grow up to about 2 $mm^3$. Imagine a tumor that is well-vascularized in early times. Such a tumor is compatible with high values for $b_{n1}$ in our model and similar to the standard FKPP with $\beta=1$ at early times followed up by $\beta=2/3$ in longer periods. It should be noted that experiencing hypoxia triggers angiogenesis and tumors barely develop angiogenesis and this result has been included for comparison.
	
	Tumors with smaller values for $b_{n1}$, grow until they become close to the limit imposed by oxygen supply, where most cells become inactive due to oxygen shortage. Such a halt in growth in spit of initial angiogenesis is known as post-vascular dormancy.
	Through a positive feedback loop, the ability to induce angiogenesis increases oxygen supply and grows larger gradually. Through this process, even previously hypoxic cells start to duplicate, resulting in a high acceleration in tumor growth rate and $\beta>1$ (see FIG. \ref{FIG4}). In other words, a hypoxic tumor experiences a high acceleration in growth rate during angiogenesis and exhibits $\beta>1$. The growth rate for such a tumor can be well below the growth rate of a well-fed tumor of the same size.  Once a tumor became well-oxygenated, in long times, it approaches a standard FKPP growth with $\beta=2/3$.   
	\begin{figure} 
		\centering 
		\includegraphics[width=0.9948\linewidth]{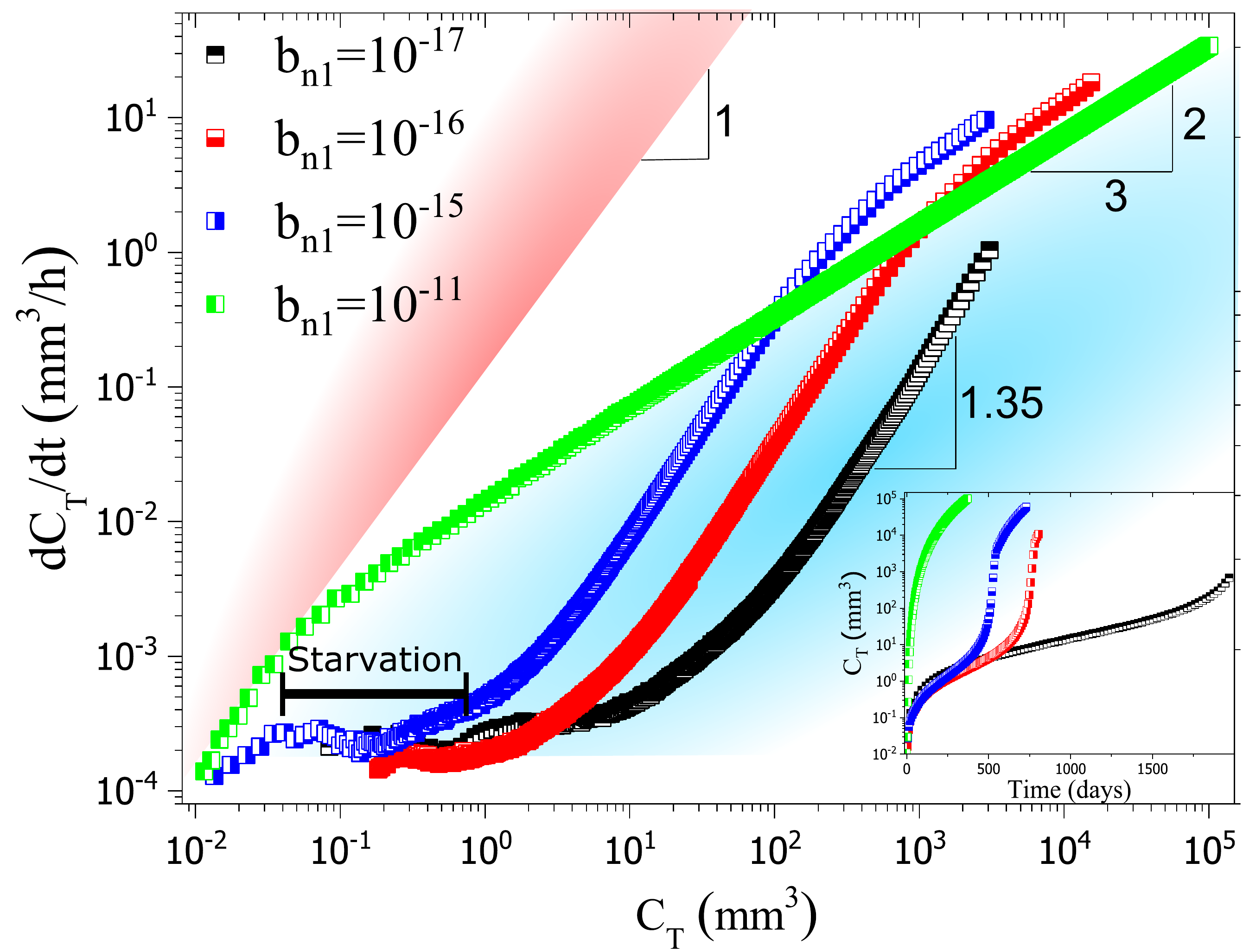}
		\caption{Growth rate versus volume for different values of $b_{n1}$ (in mol cell$^{-1}$ s$^{-1}$mm$^{-3}$). A well-fed tumor initially grows exponentially and then exhibits a surface growth with $\beta=2/3$. However, hypoxic tumors exhibit accelerated growth and $\beta>1$ during angiogenesis. The red area shows where previous models suggested having $\beta>1$ due to growing faster than exponential growth \cite{durrett2010evolutionary, Victor2020superlinear}. The blue area is where our model reveals to exhibit super-linear growth with $\beta>1$. In this area, growth is not faster than exponential growth, but the acceleration in growth due to increased oxygen supply leads to $\beta>1$. Inset: Volume versus time for different values of $b_{n1}$.}
		\label{FIG6}
	\end{figure}

	\section{Discussion} 
	Healthy tissues sustain the environment's properties, such as cellular duplication and oxygen density, by incorporating a diverse range of regulatory mechanisms. Once a group of cells avoid growth regulations and duplicate spontaneously, the rest of the activities (including oxygen regulating mechanisms) remain normal. As a result, the growing population consumes existing oxygen and soon faces oxygen shortage and that is why most tumors experience hypoxia at early stages. To grow larger, tumors need to manipulate the environment to provide more oxygen, among other alterations required for survival and progression \cite{hanahan2000hallmarks, hanahan2011hallmarks}. Tumor cells secrete chemicals and growth factors to manipulate the environment, which leads to a mutualism among tumor cells \cite{archetti2019cooperation}. While this process has been known for years, it has not been considered as the dominant and determinant factor in tumor growth\cite{turajlic2019resolving}, partially due to ample evidence on tumor cells diversity and the central role that has been assigned to evolutionary dynamics of competition among tumor cells \cite{parker2020cell, zahir2020characterizing, lakatos2020evolutionary, watson2020evolutionary}.  
	
	For a tumor that experiences hypoxia, major parts (if not all) of tumor cells, even those living on the surface, remain inactive due to oxygen shortage \cite{rodriguez2013tumor}. Respectively, the growth rate for tumors is only a fraction of its capacity, staying in dormancy \cite{hahnfeldt1999tumor, naumov2008tumor}. If the oxygen supply rate increase due to further angiogenesis, more and more cells become active, unleashing tumor capability to grow. Previously, it was suggested that the growth at such a phase might be exponential \cite{naumov2008tumor}. However, such a transition in the active fraction of tumor cells can drastically accelerate growth and increase $\beta>1$ (see FIG. \ref{FIG6}).  Thus, instead of an increase in individual duplication rates, which, as our model suggests, is too small to have a relevant effect, the transition in the number of actively duplicating cells accelerates tumor growth and leads to $\beta>1$. Respectively, the previously known Hallmark of cancer ---that is common among different cancer types--- drives super-linear dynamics among a wide range of human tumors.  The idea of accelerated growth has been under investigation in ecological literature for a long time \cite{phillips2010evolutionarily,benichou2012front, hallatschek2014acceleration}. Once a sub-population gains driver mutation, it may become dominant. However,  an accelerated growth does not lead to super-linear growth necessarily, as we showed here.     
	
	Inducing angiogenesis through mutually secreting growth factors resembles the Allee effect \cite{gatenby2019first, gatenby2020eradicating}, because it leads to an increase in individual duplication rate by providing more oxygen for tumor cells \cite{korolev2014turning}. Our results suggest that observing the super-linear growth, which is a direct result of mutualism, across human tumors is evidence for the determinant role of the Allee effect in tumor progression.  Additionally, for populations that exhibit the Allee effect, one can write $\frac{dC}{dt}=D\nabla^2C+ RC(1+C)(1-C)$ \cite{fadai2020unpacking}. An additional term as $1+C$ depicts the positive dependency of duplication rate (individual fitness) on population local density. Depending on the nature of cooperation among individuals, it is possible to substitute the local density by whole population size through substituting $C$ with $C_T$  ($C_T=\int C_t dv$). In the light of this understanding, we can interpret previously presented model, $\frac{\partial C}{\partial t}=D\nabla^2C+C(R_0+R_1 C_T) (1-C)$ \cite{Victor2020superlinear},  as an Allee effect. Thus, while the proposed model was suggested to represent an increase in fitness advantage due to driver mutations, it is compatible with the well-known Allee effect.  
	
	FIG. \ref{FIG6} revealed that super-linear growth emerges in a different parameter zone and is not necessarily faster than exponential growth. Instead, it is a result of acceleration. Such an understanding may challenge previous results in different aspects. One of the major results presented in previous work is that super-linear growth is an inherently 3D phenomenon \cite{Victor2020superlinear}. We checked if 2D environments support super-linear growth. We used a 2D version of our model to see if the acceleration in growth due to angiogenesis leads to super-linear growth. Interestingly, even in the 2D model, angiogenesis leads to super-linear growth \cite{supplemtal}. As complementary models, we used a 2D FKPP model \cite{azimzade2019effect} and a 2D individual-based model \cite{azimzade2019short} to study growth dynamics in the presence of an increasing duplication rate. All models show that super-linear growth emerges in 2D environments as well  \cite{supplemtal}. As such, it seems that mutualism (and the Allee effect) in 2D environments can lead to super-linear dynamics, suggesting that it is not an inherently 3D phenomenon.

	\subsection{Therapeutic Applications} 
	Understanding the mechanism behind super-linear growth is not only of theoretical interest. Analysis of growth rate data suggested that tumors with higher growth rate worth outcomes than those with lower rates \cite{Victor2020superlinear}. Previously, the difference between these two types (high-risk and low risk) was suggested to be initiated from differences in their evolutionary stage. Since tumors' evolutionary status is hard (if not impossible) to interfere with, there were no therapeutic implications. Our model suggests that angiogenesis is responsible for changes in growth rate and super-linear dynamics. Respectively, those tumors with higher activity are, in fact, more successful in inducing angiogenesis.  Anti-angiogenetic drugs are well developed and have been used for different cases and our findings suggest that such drugs can be effective for high-risk tumors.	 It should be noted that anti-angiogenetic drugs may fail due to genotypic changes on endothelial cells, the vascular normalization phenomenon  and the vasculogenic mimicry. More importantly, anti-angiogenic molecules' usage leads to a hypoxic tumor microenvironment, enhancing tumor invasiveness  \cite{ribatti2019limitations}. Due to such complexities, a practical intervention needs to incorporate additional interactions to predict possible failure.

	\section{Conclusion}
	Recent data revealed super-linear growth across human tumors and acceleration in growth due to the evolutionary dynamics of competition was introduced as the driving mechanism behind it. Our model suggests that while the acceleration of growth leads to super-linear growth, the emergence of cell types with higher fitness advantages does not lead to such an acceleration. Instead, hypoxic tumor cells accelerate their growth by manipulating the environment and inducing angiogenesis. This mutualism in manipulation is compatible with the Allee effect suggested being involved in tumor growth previously.
	
	\section{Acknowledgment} We would like to thank V\'{\i}ctor M. P\'erez Garc\'{\i}a for reading our draft and his fruitful comments.

 		\section{ Supplementary Material}

 	\subsection{ Agent-based exponential growth}
 	We present an agent-based model for exponential growth. In this model, we consider two scenarios as well: linear (Darwinian evolution) and punctuated evolution.
 	
 	We first have a basic model in which all cells belong to the same type with no mutations. We randomly select cells for duplication with a probability of $R$. At each time step, the total number of selections is equal to the cell number. Since there is no restriction for duplication, the whole number of cells, $N$, exponentially increases over time.   
 	
 	For Darwinian evolution, we start with 100 cells that belong to the same type. We randomly select cells for duplication. The selected cell will duplicate with a probability of $R_1$.  During each duplication, one of the daughter cells belongs to the same type and the next cell belongs to the next generation with a probability of $P_m$. We run simulations until the total number of cells reach $10^{10}$, a typical cell number in a tumor. Duplication rate for each generation increases as: $R_{i+1}=R_i (1+ \Delta R)$ with $\Delta R=0.004$ \cite{bozic2010accumulation}.
 	
\begin{figure} [h]
 		\centering
 		\includegraphics[width=0.485\linewidth]{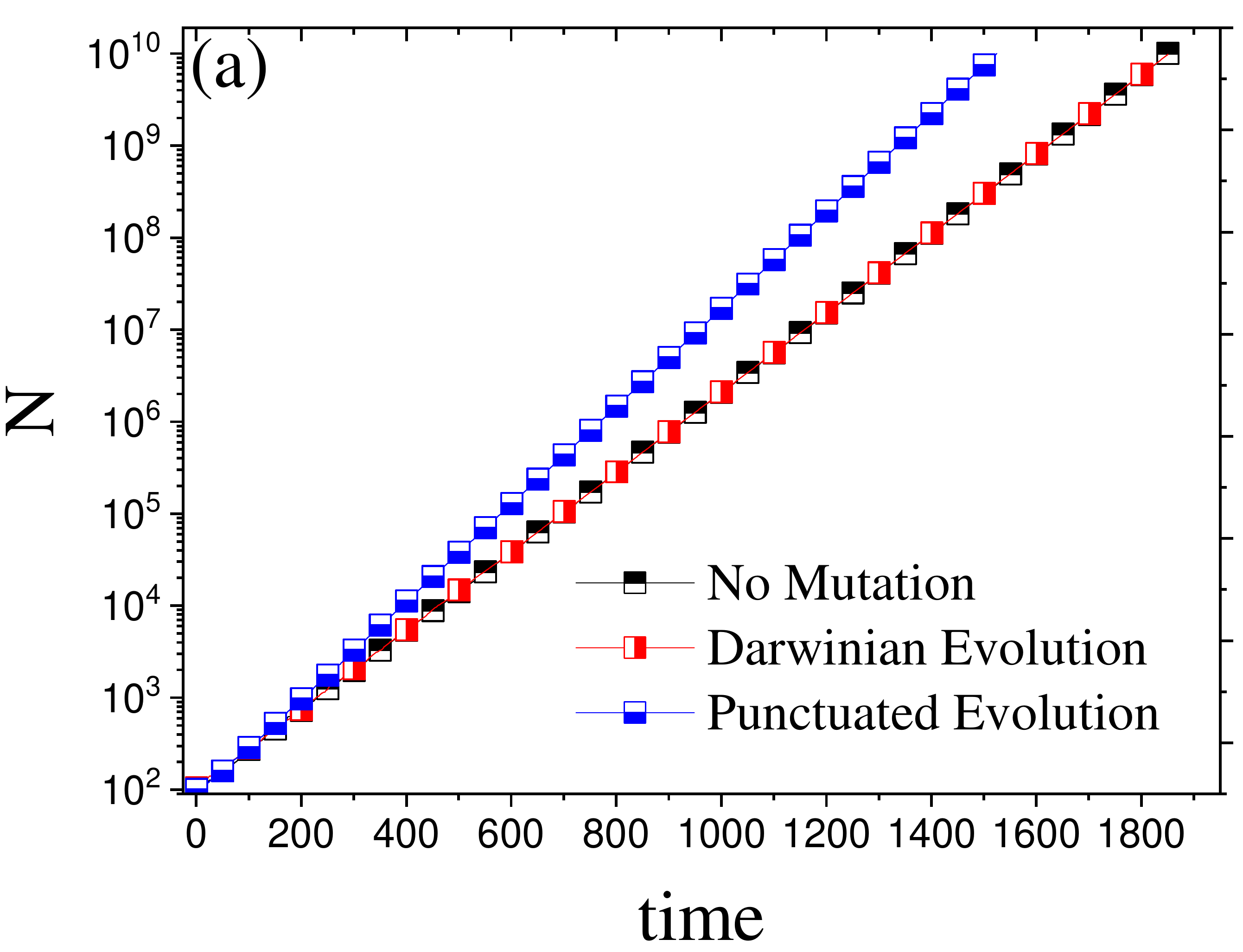} 
 		\includegraphics[width=0.485\linewidth]{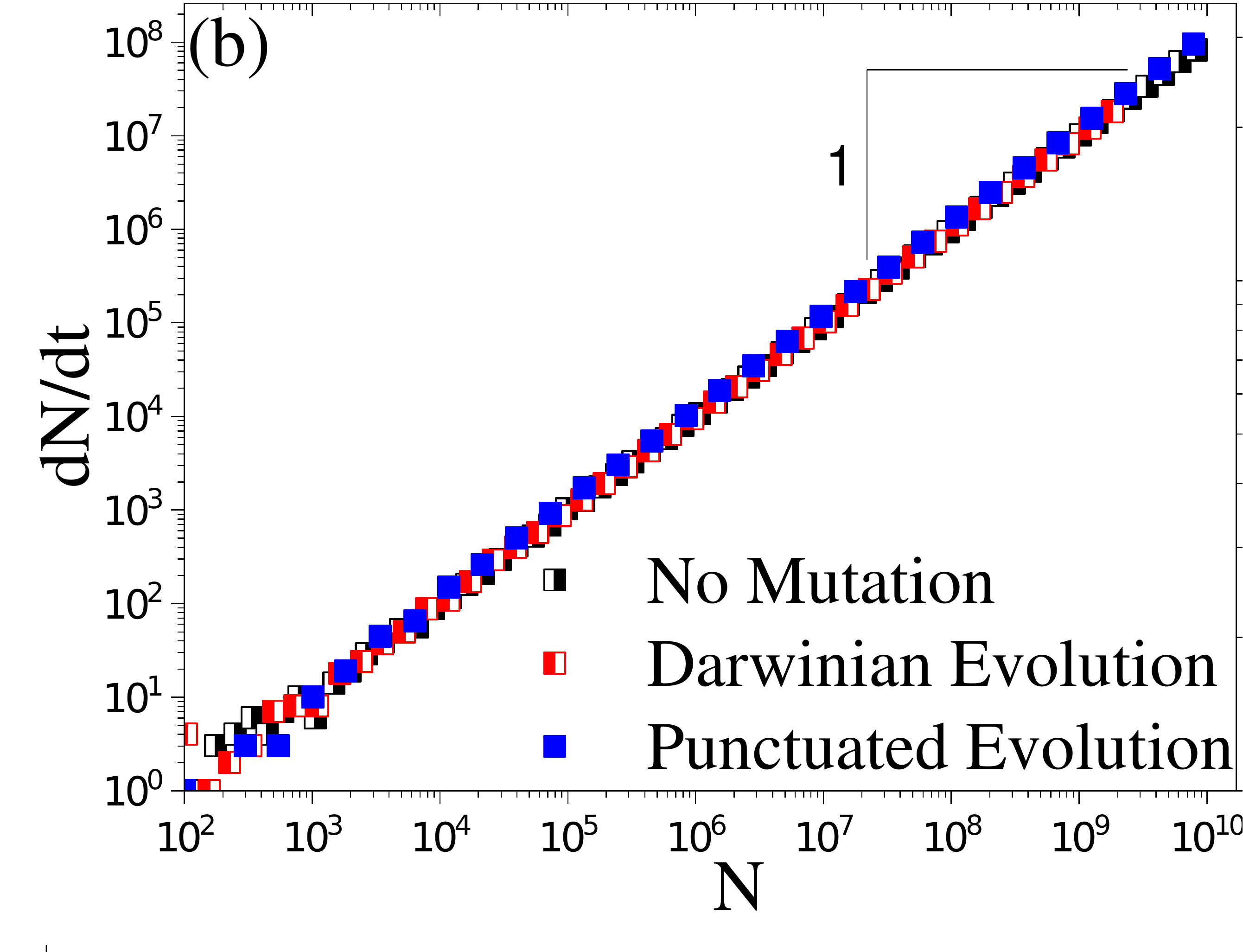}   
 		\caption{Exponential Growth.}
 		\label{FIGS1}
 	\end{figure}

 	For punctuated evolution we consider 100 cells each belonging to a different generations and while the mutation rate is zero, duplication rate increase for different generations as: $R_{i+1}=R_i (1+ \Delta R)$ with $\Delta R=0.004$. We run simulations until the total number of cells increases up to $10^{10}$, a typical cell number in a tumor with clinically relevant size.  We check for the number of cells and the number of new-born cells in each time step. As FIG \ref{FIGS1} (a) shows, cell number exponentially grows for all cases. Number of new-born cells linearly increases versus cell number (see FIG  \ref{FIGS1} (b)). As such, even in exponentially growing populations, evolutionary dynamics do not lead to super-linear growth.       
 	
 	\subsection{2D models}
 	
 	To check if 2D environments also exhibit super-linear growth, we use three different models. First, we develop a 2D version of our model. To this end, we consider the simulation box to be 2D. As the second model, we use a FKPP equation with increasing reaction term as below: 
 	
 	\begin{eqnarray}
 		\frac{\partial C } {\partial t}
 		= D_c\nabla^2 C  + (R_0 +R_1 C_T)  C (1-C)  
 		\label{Eq1}
 	\end{eqnarray} 
 	with $C_T=\int C(x,y) dx dy$. 
 	
\begin{figure} 
 		\centering
 		\includegraphics[width=0.485\linewidth]{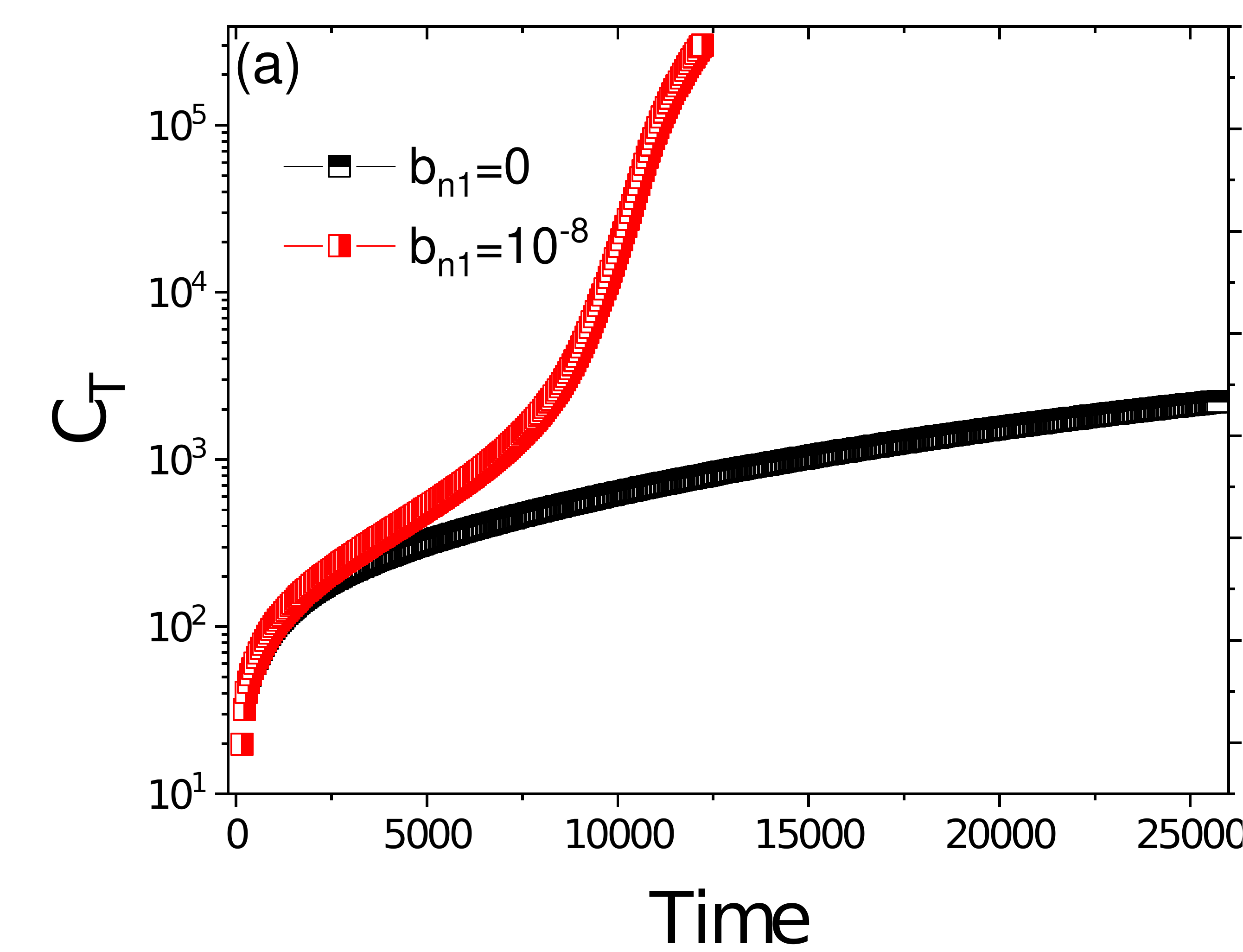} 
 		\includegraphics[width=0.485\linewidth]{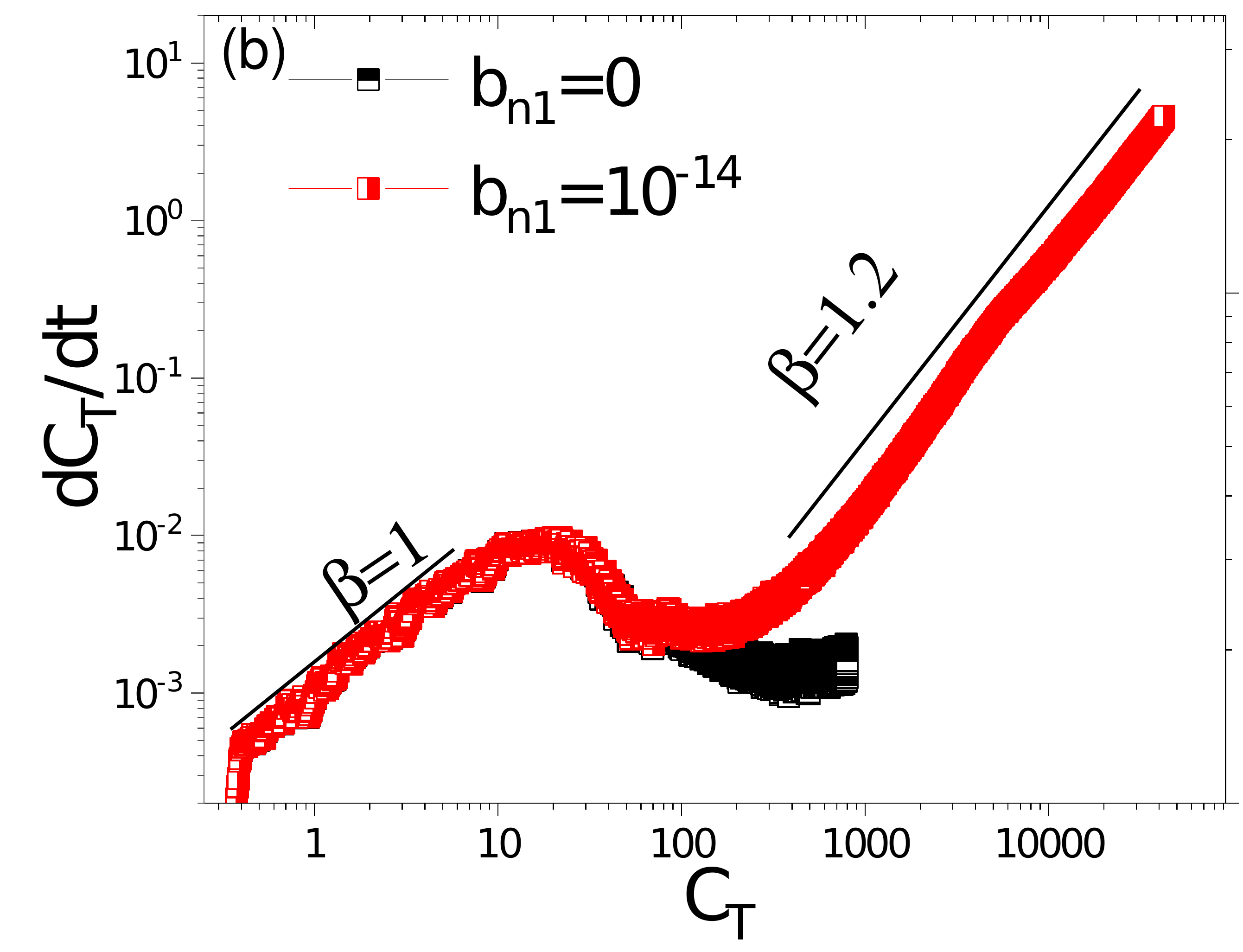}   
 		\caption{2D version of our model. Tumor volume versus time (a) and growth activity versus volume (b).}
 		\label{FIGS2}
 	\end{figure}

\begin{figure} 
 		\centering
 		\includegraphics[width=0.485\linewidth]{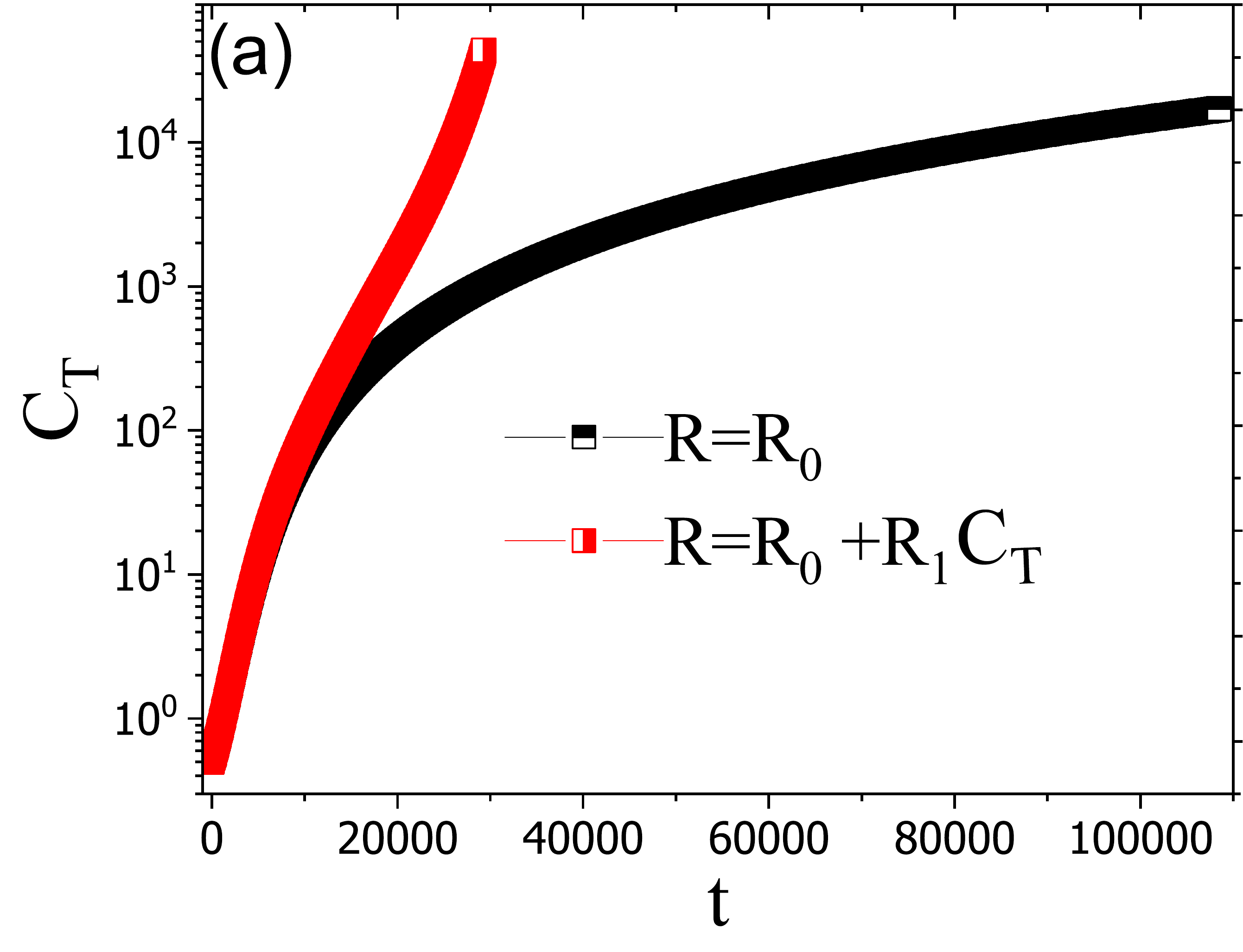} 
 		\includegraphics[width=0.485\linewidth]{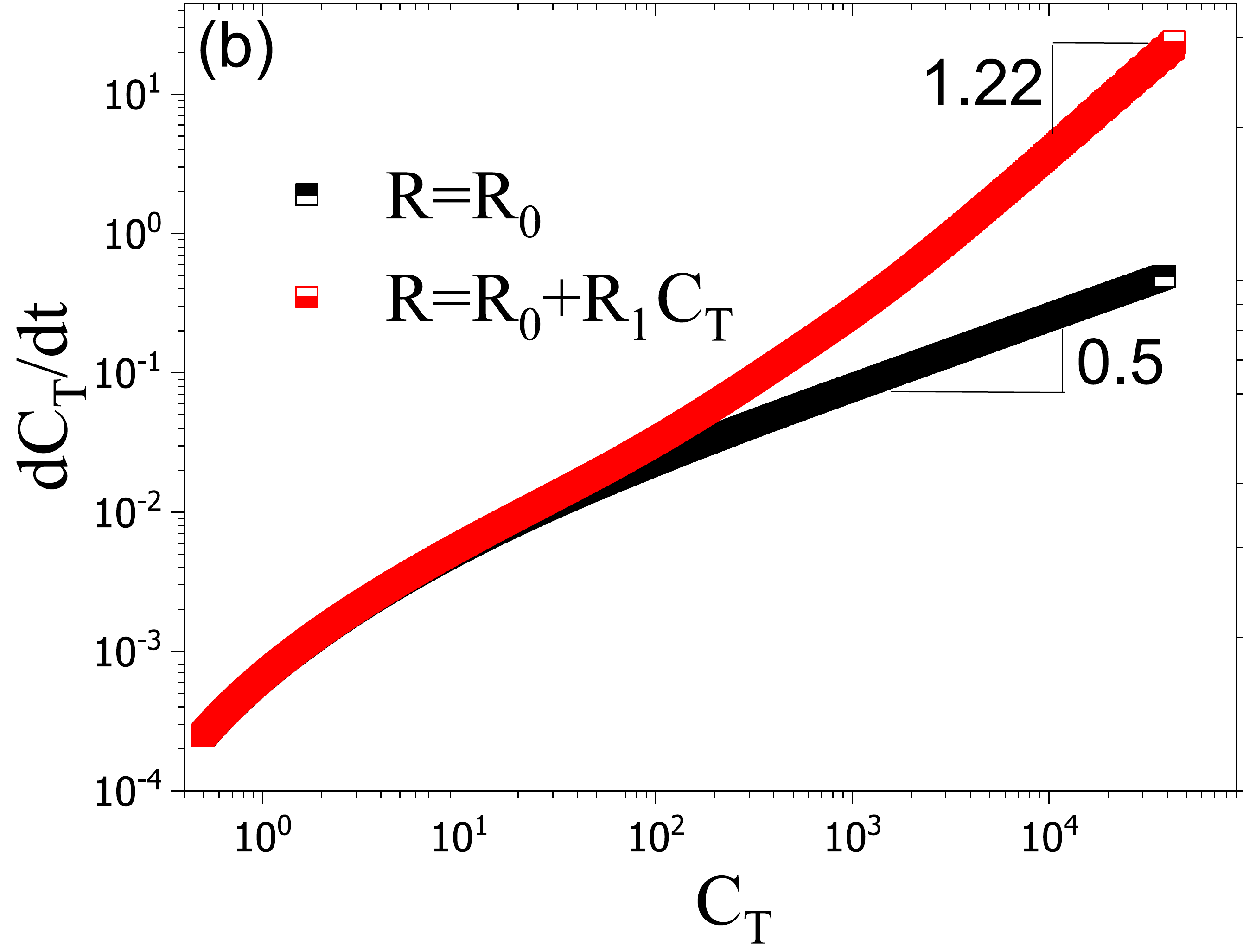}   
 		\caption{2D  FKPP.}
 		\label{FIGS3}
 	\end{figure}

\begin{figure} 
 		\centering
 		\includegraphics[width=0.485\linewidth]{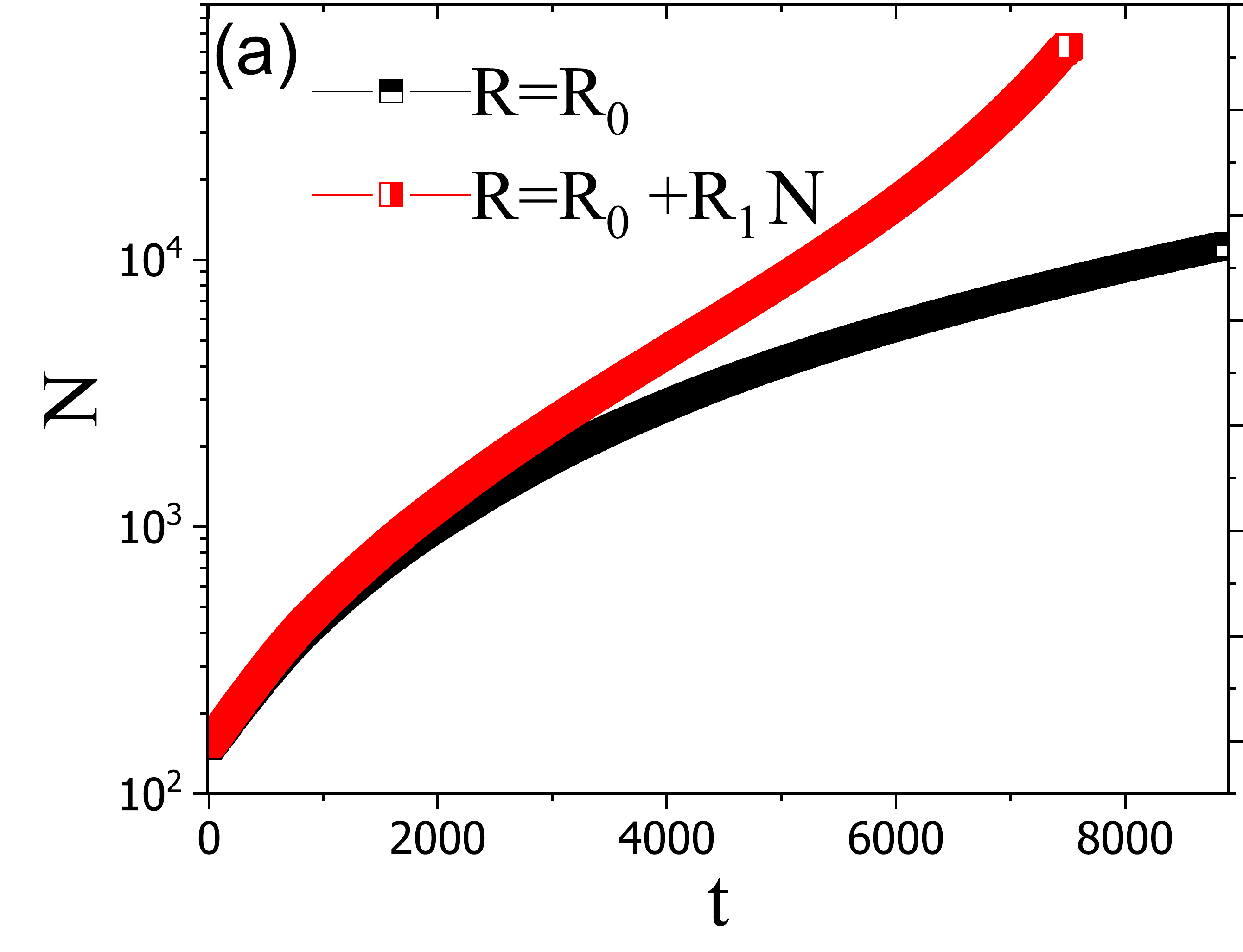} 
 		\includegraphics[width=0.485\linewidth]{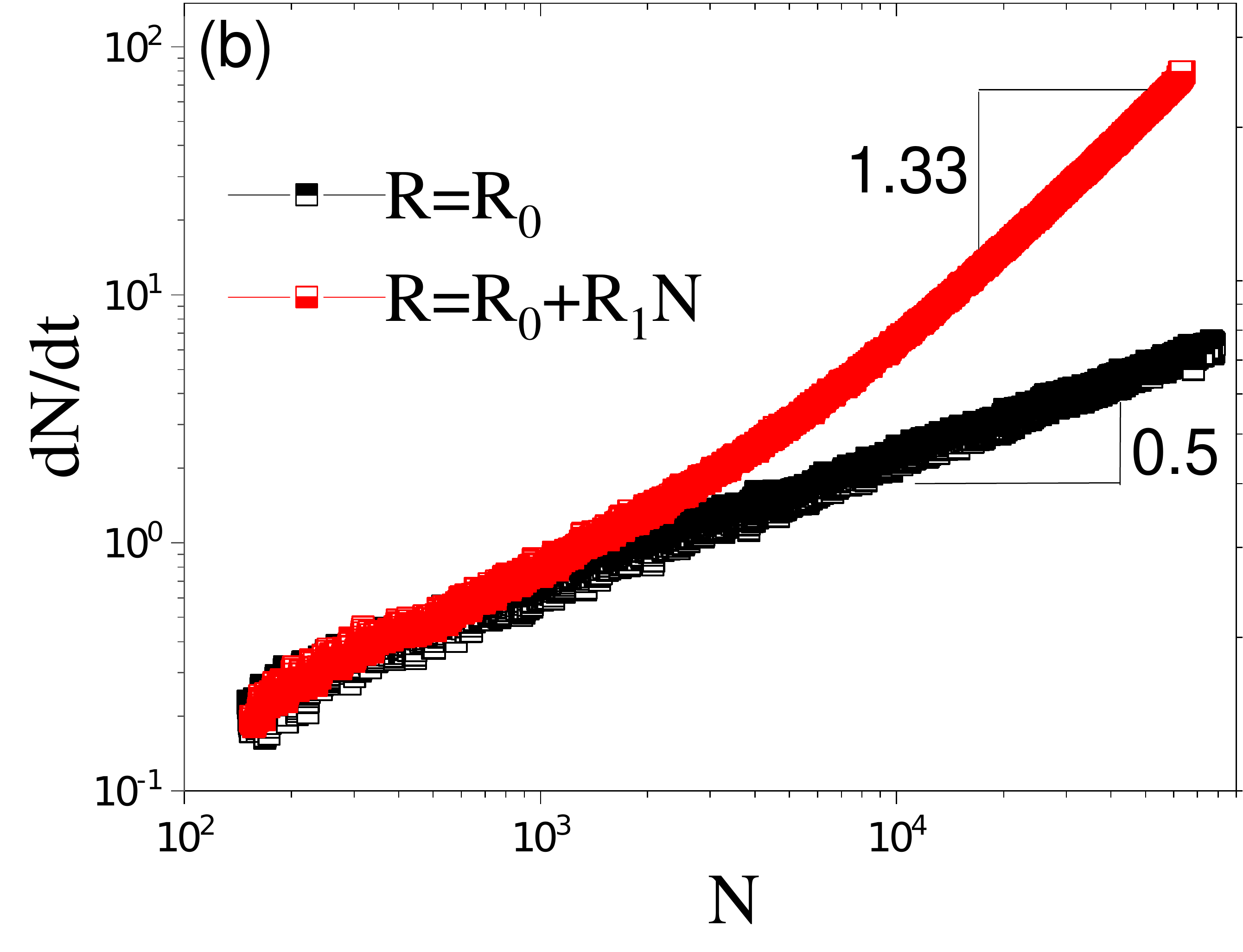}   
 		\caption{2D Discrete.}
 		\label{FIGS4}
 	\end{figure}

 	Finally we developed an agent-based model  \cite{azimzade2019short} in which instead of a fixed duplication rate of $R_0$, as tumors becomes larger, duplication rate increases as $R=R_0+R_1 N$ with $N= \Sigma C$. 
 	As the FIGs \ref{FIGS2}-\ref{FIGS4} show, all these models can lead to emergence of super-linear growth.  
 	
 	\section{Data Availability}
 	Codes for different models are provided at:   \url{https://gitlab.com/YounessAzimzade/superlinearallee}
	
\end{document}